\documentclass[prb,twocolumn,showpacs,superscriptaddress,floatfix]{revtex4}
\usepackage{graphicx}%
\usepackage{dcolumn}%
\usepackage{bm}%
\usepackage{latexsym}
\usepackage{amsmath}

\def\be{\begin{equation}}
\def\ee{\end{equation}}
\def\bea{\begin{eqnarray}}
\def\eea{\end{eqnarray}}

\usepackage{natbib}

\begin{document}

\title{Particle-hole cumulant approach for inelastic losses in x-ray spectra}

\author{J. J. Kas}
\author{J. J. Rehr}

\affiliation{Dept. of Physics, Univ. of Washington, Seattle, WA 98195-1560}

\author{J. B. Curtis}
\affiliation{Dept of Physics, Univ. of Rochester, Rochester, NY 14927}

\date{\today}

\begin{abstract}
Inelastic losses in core level x-ray spectra arise from many-body excitations,
leading to broadening and damping as well as satellite peaks in x-ray
photoemission (XPS) and x-ray absorption (XAS) spectra.
Here we present a practical approach for calculating these losses based on a
cumulant representation of the particle-hole Green's function, a  
quasi-boson approximation, and a partition of the cumulant into
extrinsic, intrinsic and interference terms. 
The intrinsic losses are calculated using real-time,
time-dependent density functional theory while the extrinsic losses are
obtained from the GW approximation of the photo-electron self-energy and
the interference terms are approximated.
These effects are included in the spectra using
a convolution with an energy dependent particle-hole spectral function. 
The approach elucidates the nature of the spectral functions in
XPS and XAS and explains the significant cancellation between extrinsic and
intrinsic losses. Edge-singularity effects in metals are also accounted for.
Illustrative results are presented for the  XPS and XAS
for both weakly and more correlated systems.

\end{abstract}

\date{\today}
\maketitle

\section{Introduction}
Inelastic losses in x-ray spectra have long been of interest.
These losses arise from electronic correlations, reflecting the
coupling of electrons and holes to excitations of the system, such as
plasmons and electron-hole pairs.  
Besides broadening and damping, 
they lead to additional features in the spectra that are not
captured by the quasi-particle approximation.  
For example, in x-ray photoemission spectra (XPS) they correspond to
satellites beyond the main quasi-particle peak and a reduction in main-peak
intensity. 
As a result, conventional
theories of x-ray spectra are usually only semi-quantitative.
Two classes of losses have been identified: i) {\it intrinsic} losses which
arise from excitations due to the sudden creation of a core hole, including
shake-up, shake-off, and  charge-transfer excitations; and ii) {\it extrinsic}
losses, which arise from similar excitations during the propagation of the
photo-electron. The extrinsic losses are often approximated in terms
of the inelastic mean free path which is related to the imaginary part of
the electron self-energy $\Sigma$.\cite{Hedin:1,MPSE}
Interference effects have also been discussed, both formally and using
approximate models.\cite{hedin99review,hmi,Campbell2002}

Surprisingly, inelastic losses in x-ray absorption spectra (XAS)
are typically smaller than
one might expect, as theoretical estimates of the intrinsic losses alone
are typically about 30\% of the main quasi-particle peak,
even in weakly corrrelated systems.\cite{hedin99review}
Moreover, losses due to satellites in XAS are almost always
neglected in practical calculations, ranging from
independent-particle to  the  Bethe-Salpeter Equation
(BSE).\cite{stobe,ocean,exciting} Their neglect at low energies
is often justified on the basis of the adiabatic approximation, 
and is often rationalized on the belief that the error is
small or only contributes a smooth background,
e.g. the many-body amplitude factor $S_0^2$ in XAS.\cite{rehr78}
The resolution of this paradox lies in the effect of the interference terms,
as both classes of losses involve similar excitations with couplings of
opposite sign and tend to cancel.\cite{hedin99review} 
While it has been argued that this
cancellation is perfect at threshold, at least for plasmon
excitations,\cite{hedin99review} we find that it is generally incomplete,
e.g., for the case of charge-transfer satellites.
Another reason for their neglect is the computational difficulty of
first principles of these losses, since various attempts
ranging from plasmon pole models,\cite{Campbell2002}
configuration interaction (CI),\cite{bagus2010}
multiplets, \cite{degrootbook} to dynamical mean-field
theory\cite{biermann12} have had only mixed success.

The aim of this work is twofold in an effort to address these issues:
first we develop a formal approach based on a
generalization of the cumulant Green's function (CG) that includes
intrinsic and extrinsic losses and interference terms; and second we
develop practical approximations for these losses which are applicable 
both to weakly-correlated and some $d$- and $f$-electron materials.
In contrast to the Dyson equation for the one-particle Green's function
$g=g^0 +g^0\Sigma g$, the CG is based on an exponential representation
in the time-domain  $g(t)=g^0(t)e^{C(t)}$, where $g^0(t)$ is the
non-interacting Green's function
and $C(t)$ is the cumulant.   This expansion is closely related to the
linked-cluster theorem and has various uses in theoretical physics.
 Its applications to spectra were significantly developed by Hedin and
collaborators,\cite{hedin99review,aryasetiawan} and
a new derivation for the one-particle Green's function
based on a functional differential equation has recently
been developed.\cite{guzzo,sky}
 While no more demanding computationally than Hedin's GW approximation for
the self-energy $\Sigma$, the CG has successfully explained satellite effects
in the XPS of weakly correlated systems,\cite{guzzo,sky,lischner,giustino} 
while the GW approximation usually overestimates the satellite position
and strength.
 Applications of cumulant methods to correlated materials based on the
quasi-boson method\cite{hedin99review} and on dynamical mean-field
theory,\cite{biermann12} have also been proposed.
  
Despite the above successes, the single-particle Green's function alone is
inadequate to describe x-ray spectra, which involves the simultaneous creation
of both a particle and a hole. Instead, our generalization here is an
analogous exponential representation of the ``particle-hole Green's function"
$G_K(t) = G_K^0(t) e^{\tilde C_K(t)}$,
where $\tilde C_K(t)$ is calculated to second order in the couplings to the
excitations in the system. The structure of $G_K$  is related to
the effective Green's function for x-ray spectra
introduced by Campbell et al.\cite{Campbell2002} (CHRB)
transformed to the time-domain (see Appendix A). 
Here $K=(c,k)$ labels the transition from a given core-level $|c\rangle$
to a photoelectron state $|k\rangle$.
A formal derivation of a related cumulant model for the
2-particle Green's function has recently been introduced by Zhou et al.\cite{sky}
The real-time representation of $G_K(t)$ considerably simplifies the
theory, and leads directly to an expression for the many-body
XAS $\mu(\omega)$ at photon energy $\hbar\omega$ as a 
convolution of the spectrum calculated in the presence of a static core
hole with an effective particle-hole spectral function $A_K(\omega)$
\begin{equation}
\label{phxas}
   \mu(\omega) =  \sum_{k} \int d\omega'\, A_{K}(\omega')
                 \mu_{K}^{0}(\omega-\omega').
\end{equation}
Here $ \mu_K^{0}(\omega)$ is the independent-particle XAS calculated
in the presence of a core-hole, and  
$A_{K}(\omega)=-(1/\pi) {\rm Im}\, G_{K}(\omega)$.
A similar convolution -- Eq.\ (49) in Ref.\ [\onlinecite{Campbell2002}] --
over the XAS fine structure $\chi_K(\omega)$ yields 
the many-body reduction factor $S_0^2$ in the XAS fine structure.
Effects of thermal vibrations and disorder can be included implicitly in
$\mu^0$ and $\chi_K$\cite{feffcr} by averaging over the structural variations.
Convolutions related to that in Eq.\ (\ref{phxas}) have also been used to
incorporate inelastic losses in the XPS photocurrent
$J_k(\omega)$.\cite{hedin99review,hmi,calandra2012}

Inelastic losses beyond the independent-particle
approximation are embedded in the cumulant
$\tilde C_K(t)$.  Partitioning the cumulant into intrinsic, extrinsic, and
interference terms then facilitates practical calculations beyond
simple models.
The factorization of the particle-hole Green's function $G_K$
implicit in the cumulant representation is analogous to that
in the classic treatment of the x-ray edge singularities by
Nozi\`eres and de Dominicis.\cite{ND}
Likewise, our cumulant treatment also accounts for edge-singularities
from low-energy particle-hole excitations in metals, as shown below.

The remainder of this paper is as follows. Sec.\ II.\ describes our
theoretical approach including the treatment of edge-singularity effects,
while Sec.\ III. contains applications to x-ray spectra for transition
metals and charge-transfer satellites.
Finally Sec.\ IV. contains a summary and conclusions.

\section{Theory}
\label{Sec:Theory}

\subsection{Particle-hole Green's function}

A detailed treatment of inelastic losses in core-level XAS
is given by Campbell et al.,\cite{Campbell2002} starting
from the formal expression
\begin{equation}
\mu \left( \omega \right) =-{\frac{1}{\pi }}{\rm Im}\,\left\langle \Psi
_{0}\left| \Delta ^{\dagger }\, \frac{1}{\omega - H + i\delta} \Delta
\right| \Psi _{0}\right\rangle .  \label{mu1}
\end{equation}
This starting point is equivalent to the many-body Fermi golden rule,
where 
$H$ is the total Hamiltonian which 
includes electron-electron interactions, 
$|\Psi _{0}\rangle $ the $N$-particle ground state of the 
system (including valence electrons and ion cores) with energy $E_0$,
and $\Delta =\Sigma _{k}\langle k|d|c\rangle c_{k}^{\dagger }c_c+hc$
is the dipole operator coupling the photon to the electronic system.
Unless otherwise specified we use atomic units,
$m=\left| e\right| =\hbar =1 $, and temperature is assumed to be zero.
The system is then partitioned into three sub-systems, a single core-level
$|c\rangle$, the valence electrons $|\Phi\rangle$, and the photoelectron 
levels $|k\rangle$; the core-hole is eliminated using a 
canonical transformation. This partition
then leads to an expression for the XAS
in terms of an effective single particle Green's function $\tilde G(\omega)$
(see Appendix A) which is a contraction of the ``particle-hole" Green's
function $G_K(\omega)$ for a discrete core state $|c\rangle$,
\begin{equation}
\label{muxas}
  \mu(\omega) = -\frac{1}{\pi} {\rm Im}\, 
   \langle c | d^{\dagger} P\, \tilde G(\omega)\, P d |c\rangle ,
\end{equation}
where $\epsilon_k = \epsilon_c + \hbar\omega$, and
$P=\Sigma_{k>k_F}|k\rangle\langle k|$ is the projection operator
onto unoccupied levels of the initial state.
As in CHRB, 
$\tilde G_K$ is approximated using a quasi-boson model Hamiltonian
in which the three subsystems
are represented in terms of a core-hole and a photoelectron coupled to
a set of bosonic  excitations, e.g., plasmons, 
particle-hole excitations, etc.,
keeping all terms to second order in the couplings. 
Next  we introduce a cumulant ansatz for $\tilde G_K$
\begin{equation}
\label{phcumulant}
  \tilde G_K(t) =  \tilde G_K^0(t) e^{\tilde C_K(t)}, 
\end{equation}
where $\tilde G_K^0(t) =  g^0_c(t) g^0_k(t),$ and
$g^0_c(t)$ and $g^0_k(t)$ are the bare core-hole and photoelectron
Green's functions, respectively, 
the latter being calculated in the presence of the core-hole.
 The generalized cumulant $\tilde C_K(t)$ is
determined (Appendix A) by transforming
Eq.\ (32) of CHRB to the time-domain, and matching the leading terms in
powers of the quasi-boson coupling constants,
\begin{equation}
\label{tilde-g}
 \tilde C_K(t) = \int d\omega\, \gamma_K(\omega)
 (e^{i\omega t} - i\omega t -1).
\end{equation}
This ansatz is similar to that derived by Zhou et al.,\cite{sky} where
$G_K$ is the 2-particle Green's function of the Bethe-Salpeter
equation. However, the cumulants differ in technical details.
The Landau representation\cite{landau44} of Eq.\ (\ref{tilde-g})
ensures that the particle-hole spectral function
\begin{equation}
\tilde A_K(\omega) = -\frac{1}{\pi} {\rm Im} \int dt\, e^{i\omega t}\,
\tilde G_K^0(t)e^{\tilde C_K(t)}
\end{equation}
remains normalized with an invariant centroid. Thus the effect
of the bosonic excitations is a transfer of spectral
weight from the main peak to the satellites while the overall strength is
conserved. Note that lifetime broadening due to the photoelectron
interactions is included naturally, while core-hole lifetime effects
can be treated by adding a damping term, $-\Gamma_c |t|,$ to the cumulant. 
In addition to describing the excitation spectrum, the cumulant formalism
simplifies the calculation of both the quasi-particle peak shift
(or relaxation energy) $\Delta_E$ and
the net quasiparticle weight (or renormalization constant) $Z_K$ in terms of
the kernel $\gamma_K(\omega)=\beta_K(\omega)/\omega^2$,\cite{hedin99review}
\begin{eqnarray}
\label{eqn:qpZfactor}
Z_K &=& e^{-a_K}, \\
a_K &=&  \int \frac{\beta_K(\omega)}{\omega^2}d\omega ,  \\
\Delta_E &=& \int_{0}^{\infty}\frac{\beta_K(\omega)}{\omega} d\omega.       
\end{eqnarray}
The excitation spectrum 
$\beta_K(\omega)$ for XAS is implicit in the 
particle-hole cumulant $C_K(t)$, and hence the structure of
$\gamma_K(\omega)$. This structure can be
understood formally in terms of
the {\it fluctuation potentials} or oscillator strength amplitudes
$V^q$ that couple electron and hole states to boson excitations
$q$ with energies $\omega_q$ \cite{Campbell2002,hedin99review,hmi,bard85}
(see Appendix A). Formally the fluctuation potentials can be obtained by 
diagonalizing the screened coulomb potential $W=\epsilon^{-1}v$.
As an illustrative example, the fluctuation potential for plasmons 
of momentum $q$ in the homogeneous electron gas is
$V^q = [v_q\omega_p^2/2\omega_q)]^{1/2} \exp(i {\bf q}\cdot {\bf r})$,
where $v_q=4\pi/q^2$ is the bare Coulomb interaction.
If recoil due to plasmon disperson is ignored,
$\gamma_K(\omega)$ can be expressed as a perfect square,
\begin{equation}
\label{perfectsquare}
\gamma_K(\omega) = \sum_q \left |V_{\bf k k+q}^q g^{0}_{\bf k+q}(\omega - \omega_{q}) -
\frac{ V^q_{cc} }{\omega_q}\right|^2 \delta(\omega-\omega_q).
\end{equation}
This representation is similar to that in the treatment of inelastic losses in
XPS,\cite{hedin99review,hmi,bard85} where the fluctuation potentials
$V^q$ are discussed in detail.  Unfortunately, this form does not seem
to be computationally useful except in simple models, due to
the non-local character of the interference terms. Moreover, its
general validity is questionable.  Thus instead, we partition $\gamma_K$,
and hence $C_K(t)$, into intrinsic $(c)$, extrinsic $(k)$, and
interference terms $(kc)$, respectively, i.e.,
\begin{eqnarray}
\tilde \gamma_K(\omega) &=& \gamma_{c}(\omega) + \gamma_{k}(\omega) +
\gamma_{ck}(\omega), \\
\tilde C_K(t) &=& C_{c}(t) + C_{k}(t) + C_{ck}.
\end{eqnarray}
The amplitudes $a_K$ and shifts $\Delta_E$ can also be split into intrinsic,
extrinsic, and interference contributions. 
The intrinsic and extrinsic parts  of $\Delta_E$ are
formally equivalent to those of the GW approximation, while the
interference term tends to reduce the shift. Similarly, the
renormalization constant can be related to derivatives of the
self-energy at the quasiparticle energies, i.e., $Z_{k} = \exp(-a_{k})$
where $a_{k} = d\Sigma_{k}/d\omega + d\Sigma_{c}/d\omega - a_{ck}$,
and again the interference terms reduce many-body effects, restoring
weight to the quasiparticle peak.  Note that in using a final
state rule (or static BSE) approximation as the starting point in
Eq.~(\ref{phxas}), this shift is already taken into account. In
order to avoid double counting and considering the approximations used
for the interference terms, we subtract this shift in our final results.

If interference is neglected, the particle-hole Green's function
would be simply a product
of the core-hole Green's function $g_c(t) = g_c^0(t) e^{C_{c}(t)}$
and the damped final state Green's function in the presence of
a core hole ${\tilde g}_k(t) = \tilde g^{0}_k(t) e^{C_{k}(t)}$.
This approximation implies that the intrinsic and extrinsic losses
are independent and additive. However  this yields XAS satellite strengths
that are generally too large.  Thus the interference terms are usually
essential; they provide an
energy dependence which tends to cancel the extrinsic and intrinsic
losses near threshold, due to the opposite signs of the hole and photoelectron charges,
while at very high energies only the intrinsic losses remain.
This phenomena is characterized as an adiabatic to sudden transition.
It is used to justify the adiabatic approximation 
and the usual neglect of inelastic losses near threshold, i.e., 
well below the characteristic excitation energy $\omega_p$.

In any case, the above partition of the cumulant $\tilde C_K(t)$ 
permits independent treatments of the various terms. This
is advantageous computationally as the physics of the intrinsic
and extrinsic losses can differ significantly.
Here we treat the intrinsic losses with the cumulant $C_c(t)$ for
the core-hole Green's function using a real-space, real-time
method of Kas et al. (KVRC),\cite{KRC} as described below.  This local approach
was found to account well for charge-transfer satellites in the XPS
of transition metal oxides.
In contrast, the extrinsic losses are treated using the cumulant approximation
$C_k(t)$ for the photoelectron Green's function. Finally 
the interference terms $C_{ck}(t)$ are approximated, e.g.,
 with an interpolation formula.
We note, however, that the partition of the cumulant is somewhat arbitrary,
and can be tailored for computational purposes.  
For example, CHRB lump the quasi-particle part of the cumulant into the
damped particle Green's function in the presence of a core hole,
$\tilde g_k(t)= {g'_k}^{0}(t)\exp[C_k^{qp}(t)]$, so that the net spectral function
only contains the satellite contributions. The full many body XAS
$\mu(\omega)$ can then be expressed as a convolution of an independent
particle XAS with a spectral function as in Eq.\ (\ref{phxas}),
where $\mu^{0}(\omega)$ is the independent particle XAS calculated
in the presence of a core hole.

\subsection{Intrinsic Losses}

The intrinsic losses are given by 
the leading factors in $G_K(t)$, which correspond to a
cumulant representation of the core-hole Green's function,
\begin{equation}
g_c(t) = i\theta(t) e^{-i \epsilon_c t + C_c(t)},
\end{equation}
where $\theta(t)$ is the unit step function.
In terms of the fluctuation potentials,
the intrinsic excitation spectrum $\gamma_c(\omega) \equiv
\beta_c(\omega)/\omega^2$ can be expressed as
$\beta_c(\omega) =\sum_q |V^q_{cc}|^2 \delta(\omega-\omega_q)$.
Physically $V^q/\omega_q$ can be interpreted as a shake-up amplitude,
and from first order perturbation theory, is equivalent to a many-body overlap 
integral between the ground state and the shake-up excited state
$|K {q}\rangle$ with a boson in state ${q}$ and a core hole in
level $|c\rangle$.\cite{Campbell2002}

The localized nature of a deep core-hole in x-ray spectra has
led us to consider a real-space, real-time approach
for the intrinsic cumulant $C_c(t)$ introduced by
KVRC,\cite{KRC} which is not limited to small clusters.
Our treatment is based on a time-dependent density functional theory
formalism (RT-TDDFT) inspired by that of Bertsch and Yabana \cite{yabana96}
for optical response.  Such methods are advantageous
for calculations of density response, since they are quantitative yet
require little computational effort beyond successive applications of
Kohn-Sham DFT in the time-evolution of the system.
RT-TDDFT has been successfully applied both to linear and
non-linear optical
response.\cite{yabana2006,takimoto2007,vila2010,otobe2009}  The approach
has also been applied to calculate the quasiparticle contribution
$\mu^{qp}(\omega)$ in Eq.\ (1) to core-level x-ray absorption spectra, using a time-correlation approach that ignores satellites.\cite{ajlee}

We will refer to $C_c(t)$ as the {\it Langreth cumulant},\cite{langreth70}
since a similar formalism was introduced to calculate edge singularities in electron gas models of deep-core x-ray spectra, following the classic treatment
of Nozi\`{e}res and de Dominicis.\cite{ND}  Transforming Langreth's
approach \cite{langreth70,bard85} to real-space, $C_c(t)$ can
be approximated to second order in the core-hole potential by
\begin{align}
  C_c(t) &=  \int d\omega\,\beta_c(\omega)
\frac{e^{i\omega t} -i \omega t -1}{\omega^{2}}, \label{eq:cum}\\
  \beta_c(\omega) &= \int d^3 r d^3 r'\, V^*({\bm{r}}) V({\bm{r'}})\,
{\rm Im}[\chi(\bm{r},\bm{r'},\omega)]. \label{eq:betac}
\end{align}
The time dependence $[\exp(i\omega t)-i\omega t -1]/\omega^{2}$
arises from the transient core-hole potential, which
turns on at time $t=0$  and then off at time $t$. 
Here $V({\bm r})$ is the bare core-hole potential, and the response function,
\begin{equation}
\chi(\bm{r},\bm{r'},\omega) = i \int dt\, e^{i \omega t}\langle
\hat\rho({\bm{r}},t)\hat\rho({\bm{r'}},0)\rangle,
\end{equation}
is equivalent to the dynamic structure factor
which is directly related to the local density-density correlation function
$\langle\hat\rho({\bm{r}},t)\hat\rho({\bm{r'}},0)\rangle$,
where $\hat\rho({{\bf r},t})$ is the density operator. 


 In more detail, our approach is as follows: $\beta_c(\omega)$ is obtained from
the Fourier transform of the ``core-response" function
 $\Delta_c(t)$ using the relations\cite{KRC}
\begin{eqnarray}
\label{eqn:betaresponse}
\beta_c(\omega) &=&  \omega\, \textrm{Re} \int dt e^{-i\omega t}\Delta_c(t), \\
\Delta_c(t) &=& \int d^3 r V(\vec r) \delta \rho(\vec r,t),
\end{eqnarray}
where $\delta \rho(\vec r,t)$ is the change in electron density from
equilibrium due to the core-hole perturbation, and $V(\vec{r})$ is the potential
due to the presence of the core-hole.\cite{KRC} This function
$\Delta_c(t)$ is computed using RT-TDDFT via a modified version of the
SIESTA framework.\cite{siesta}   

\begin{figure}[h] 
\label{fig:c60core}
\includegraphics[height=0.825\columnwidth, angle=-90]{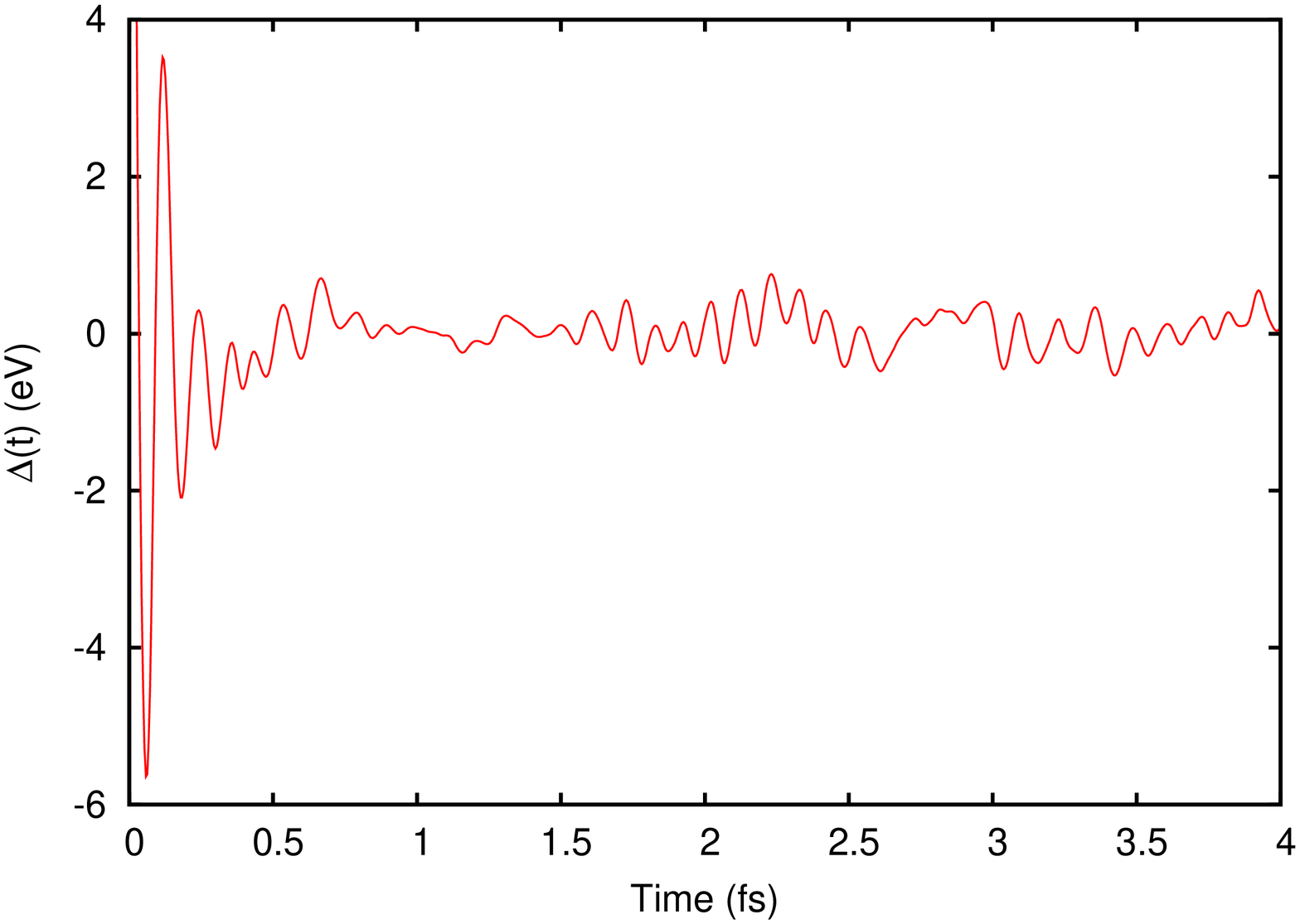}\vspace{3 mm}
\includegraphics[width=0.80\columnwidth]{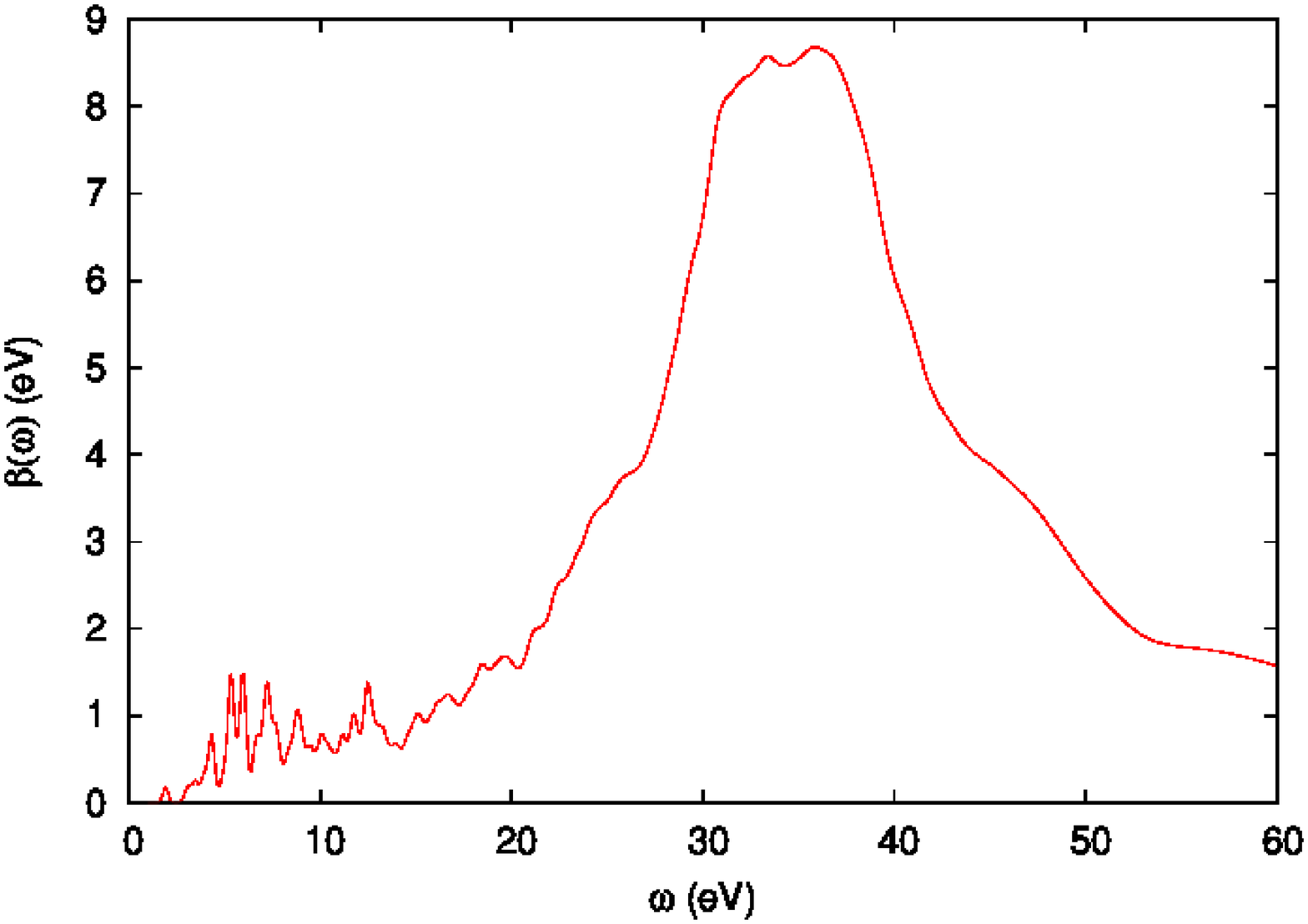}\vspace{3 mm}
\includegraphics[width=0.80\columnwidth]{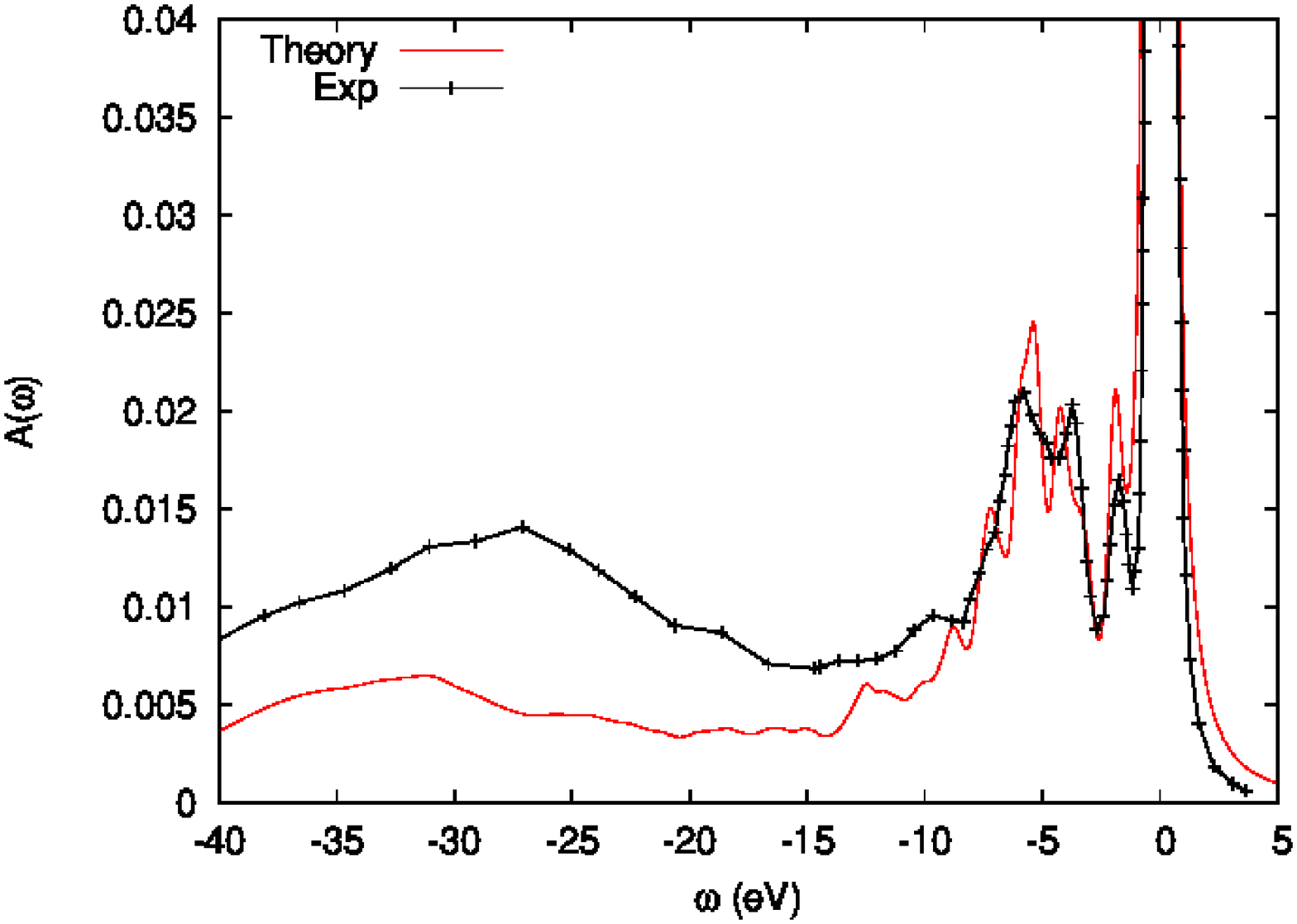}\vspace{3 mm}
\caption{(Color online) 
a) (top) The core response function $\Delta(t)$;
b) (middle) $\beta_c(\omega)$ for $\textrm{C}_{60} $,
and c) (bottom) Theoretical core-level spectral function of $\textrm{C}_{60}$
compared to experimental $1$s XPS.\cite{Leiro2003}  }
\end{figure}
The normalized core-hole spectral function $A_c(\omega)$, which
characterizes the
distribution of intrinsic excitations, is then obtained from the
imaginary part of $g_c(\omega)$ in frequency space,
\begin{equation}
\label{eq:spfcngf}
  A_{c}(\omega) = -\frac{1}{\pi}{\rm Im}\int dt\, e^{i\omega t} g_c(t).   
\end{equation}

This procedure is illustrated here for
the fullerene molecule $\textrm{C}_{60}$ in
the gas phase, and results for $\Delta(t)$ and $\beta_c(\omega)$  are
shown (top and middle respectively)
in Fig.\ 1, along with a comparison of the spectral function with
experimental XPS results. Additional details for the case of $C_{60}$ will be published
elsewhere.\cite{Shirleyetal16}
As an illustration of the theory, we present calculations of
the core-hole spectral function
$A_{c}(\omega)$. As discussed below, $A_c$ is closely related to the
 core XPS, and is in good agreement with the experimental spectrum,
 apart from a smooth background term, and a mismatch of the high binding energy
peak in the experiment at about $-27$ eV, and in the theory at about $-31$
eV. This energy shift could be due to a variety of factors including
the use of a local basis set or the approximate kernel used in the
TDDFT. For this example we have not included extrinsic and interference
effects, which are expected to increase the weight of the plasmon peak
relative to the lower energy peaks.

As discussed in Ref.\ [\onlinecite{hedin99review}],
the XPS photocurrent $J_k(\omega)$ can be approximated by $A_c(\omega)$
when the energy dependence of the matrix elements can be neglected,
apart from a smooth background,
\begin{equation}
\label{photocurrent}
J_k(\omega) = |M_{ck}|^2 A_{c}(\omega) \approx A_c(\omega),
\end{equation}
where $M_{ck}$ is the dipole transition element.  The
extrinsic and interference terms may also be
important, although in monoatomic weakly correlated systems, they
mostly affect size of the satellites and not their
shape.\cite{hmi,guzzo} For quantitative calculations, a particle-hole spectral 
function  $\tilde A_K(\omega)$ tailored for the XPS photocurrent 
is needed. This spectral function differs from that in XAS due
to the differences in boundary conditions, such as the effects
of the surface on the fluctuation potentials. In particular, in XPS,
the finite inelastic mean free path of the photoelectron limits the depth from
which electrons can reach the surface (and detector). Photoelectrons
originating deeper in the material, i.e., with longer mean free paths, 
have larger probabilities of creating excitations. On the other hand,
small mean free paths are associated with larger couplings, and thus
the two effects compete. 
For a deep hole coupled to ideal plasmons or bosons,
the cumulant representation of $g_c(t)$ given by Eq.~(\ref{eq:cum})
and (\ref{eq:betac}) is exact.\cite{langreth70}
It can be shown that this approximation is equivalent to the
decoupling approximation of
Ref.\ [\onlinecite{guzzo}]. However, corrections to the
2nd order approximation for the cumulant will generally affect the
structure of higher order satellites. 


\subsection {Edge singularities}
\begin{figure} [b]
  \label{fig:Al_beta}
  \includegraphics[height=0.8\columnwidth, angle=-90]{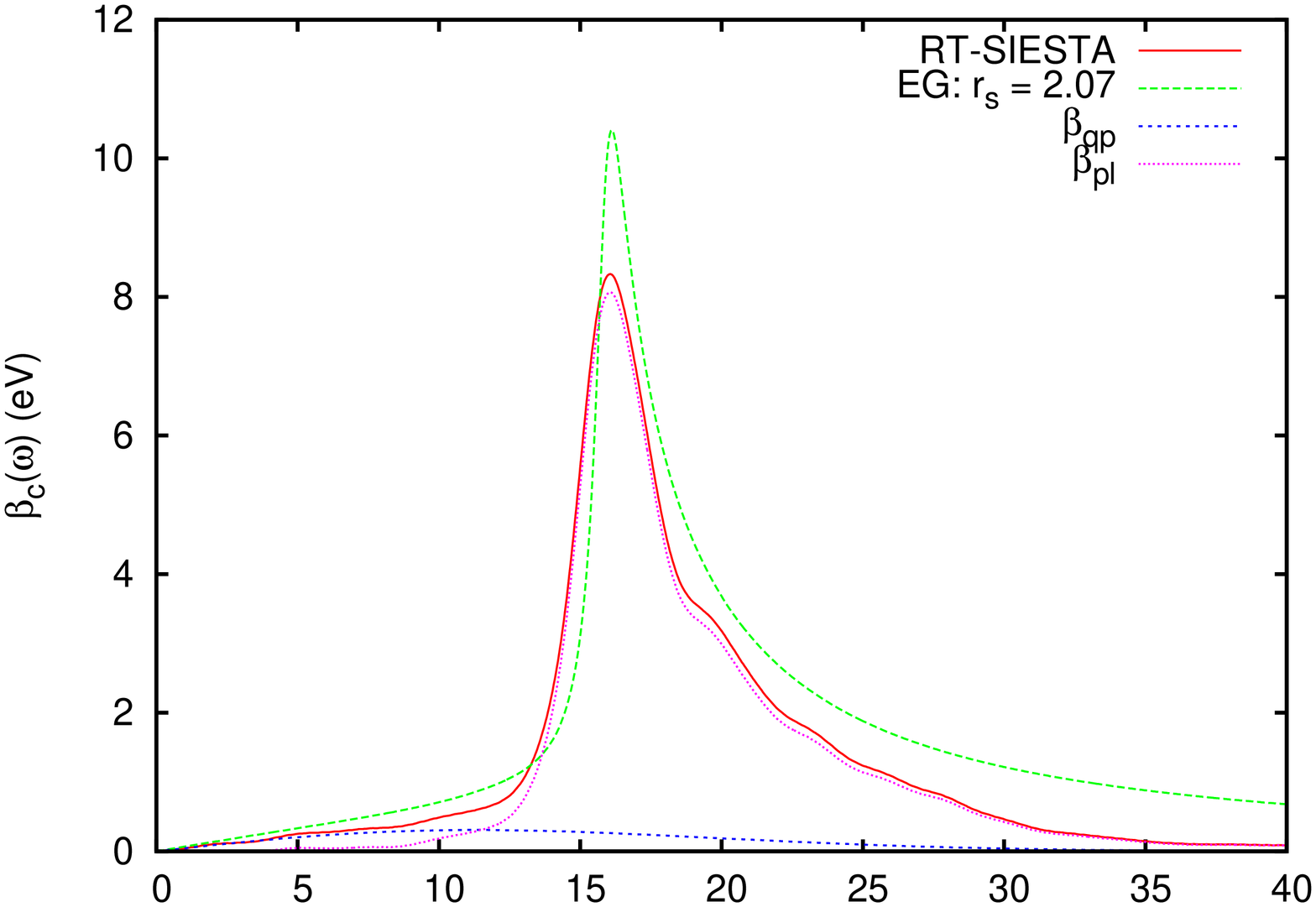}\vspace{3 mm}
  \includegraphics[height=0.8\columnwidth, angle=-90]{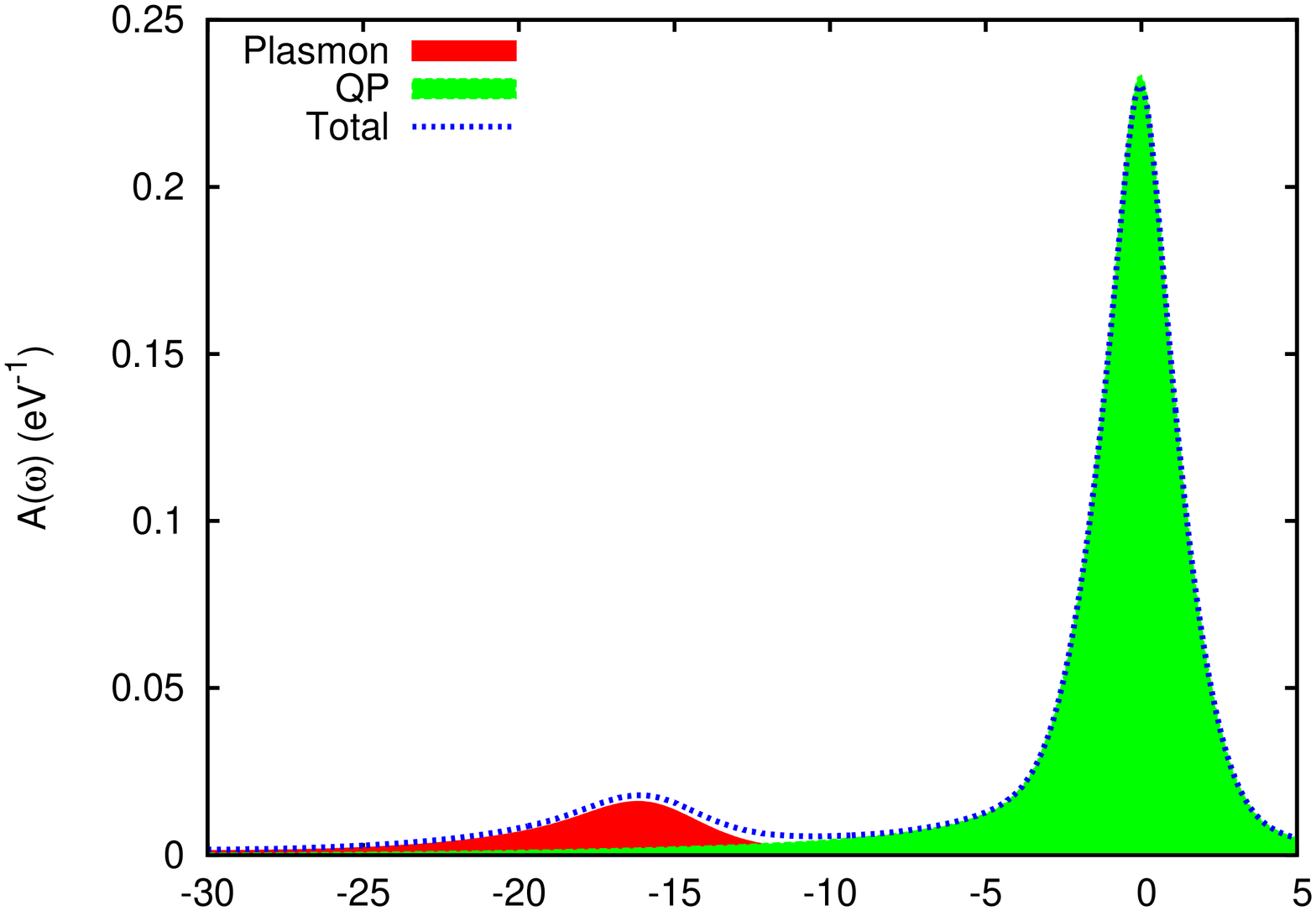}\vspace{3 mm}
  \caption{Results of the calculated core excitation spectrum (top) of Al using the real-time TDDFT method compared to that of an electron gas with $r_s=2.07$. The model particle hole part and plasmon parts are also shown. The bottom portion shows the total spectral function (dashed) along with partition into the main (green filled) and satellite (red filled) parts.  } 
\end{figure}

For metallic systems, edge singularities have been predicted 
due to existence of particle-hole excitations at zero
frequency.\cite{ND,langreth70}
In this case $\beta_c(\omega) \sim \omega$ at low frequency,
and the usual description of the
quasiparticle peak becomes problematical, because the quasiparticle weight
is strictly zero, a phenomenon known as the Anderson orthogonality
catastrophe.\cite{anderson67} 
Instead the spectral function, and hence the threshold peak in XPS
is predicted to have a power-law singularity
$A_c(\omega) \sim \omega^{\alpha-1}$, as discussed in many
works.\cite{anderson67,ND,langreth70,doniach70}
Anderson, and later Nozieres and de Dominicis derived the
exponent $\alpha = 2\Sigma_l (2l+1) (\delta_l/\pi)^2 $ in terms of
the phase shifts $\delta_{l}$ at the Fermi energy induced by
the core-hole potential. Subsequently Langreth related $\alpha$
to the low frequency behavior of the loss function,
\begin{equation}
  \alpha = \sum_{q <2 q_{F}}\frac{|{\upsilon}_{q}|^{2}N(\epsilon_F)}{q v_{F}
    |\epsilon({\bf q},0)|^{2}}, 
\end{equation}
where $N(\epsilon_F)$ is the density of states at the Fermi level, $v_{F}$ is
the Fermi velocity, and $\upsilon_{q}$ is the Fourier transform of the
core-hole potential to momentum space. This power-law singularity
corresponds to a logarithmic behavior of the cumulant in the long time
limit $C(t) \rightarrow \ln(t^{-\alpha})$.

Here we have reinvestigated this singular behavior within
the cumulant
formalism. We generalize the quasi-particle
line-shape to an asymmetric form that includes a power-law singularity,
analogous to that of Doniach and Sunjic.\cite{doniach70}
To do this we further partition the core-hole excitation spectrum 
$\beta_c(\omega)$ into ``particle-hole'' and ``plasmon''
contributions, where the particle-hole contributions only contribute at
low frequencies and give the part of the excitation
spectrum $\beta_{ph}(\omega)$ which is
linear in frequency. This linear part 
responsible for the logarithmic divergence of the cumulant
in the long time limit, and hence the singular behavior at threshold.
We therefore redefine the main peak in terms of the normalized spectral
function $A_{ph}$ resulting from the ``particle-hole'' contribution of the excitation spectrum
$\beta_{ph}(\omega)$, which is
defined with an {\it ad hoc} exponential damping factor to enforce
normalizability of the spectral function
\begin{equation}
  \beta_{ph}(\omega) = \alpha \omega e^{-\omega/\omega_{p}}.
\end{equation}
The exponent $\alpha$ corresponds to the linear coefficient at low frequencies
 and $\omega_{p}$ is the plasmon frequency. The precise nature of the damping
is not important as it does not effect the edge singularity, and the net excitation spectrum is conserved by setting 
the plasmon part as $\beta_{pl}(\omega) = \beta(\omega)-\beta_{ph}(\omega)$.
Substituting this form into Eq.~(\ref{eq:cum}) for the particle-hole cumulant
gives
\begin{equation}
  C_{ph}(t) = -i \alpha \omega_{p} t - \alpha \ln(1 - i \omega_{p} t).
\end{equation}
The main peak of the core-hole Green's function and the associated
spectral function are then
\begin{eqnarray}
  g_{ph}(t) &=& -i e^{-i (\epsilon_{c}+\alpha \omega_{p}) t -\alpha\ln(1 -
      i\omega_{p} t)}, \\
  A_{ph}(\omega) &=& e^{-a_{pl}}\frac{e^{-\tilde{\omega}/\omega_{p}}}{\Gamma(\alpha)}\frac{\omega_{p}^{-\alpha}}{\tilde{\omega}^{1-\alpha}},
\end{eqnarray}
where $\tilde{\omega} = \omega - \epsilon_{c} - \alpha \omega_{p}$ is
the frequency relative to theshold.
This representation has the correct behavior at long times as well as 
at $t = 0$, where the cumulant must vanish to preserve normalization.
The weight of the quasiparticle spectral function, i.e., the generalized
renormalization constant) is given by the plasmon-part
$Z = e^{-a_{pl}}$,
which is reduced from unity due to the high energy (e.g.  plasmon)
excitations,
\begin{align}
  a_{pl} &= \int d\omega \frac{\beta_{pl}(\omega)}{\omega^{2}},
  \nonumber \\
  \beta_{pl}(\omega) &= \beta(\omega) - \beta_{ph}(\omega).
\end{align}
As an example, we show the separation of the particle-hole and plasmon
peaks for fcc Al in Fig.\ 2 (bottom).

We note that this edge-singularity correction only appears in 
the intrinsic spectral function in metals; there is generally no contribution
from the extrinsic losses for photoelectron states  $k$ far above threshold.
However, for XAS an additional Mahan edge singularity factor
\begin{equation}
\label{mahan}
\tilde\mu(\omega) \sim \omega^{-2\delta_l/\pi},
\end{equation}
appears in the dipole-matrix elements 
due to the non-orthogonality of the one-particle levels with and without the
core hole.\cite{Campbell2002,ND}
Even in insulators, one may expect a non-singular enhancement factor
given by Eq.\ (\ref{mahan}), with the threshold Fermi
energy set at mid-gap.  In contrast the main peak in the extrinsic spectral
function $A_k(\omega)$ has an asymmetric Fano line shape,
as discussed by Aryasetiawan et al.\cite{aryasetiawan}
Finally, we note that the finite lifetimes of the core-hole and
photoelectron will
broaden the observed edge singularity.

\begin{figure}[t]
  \label{fig:CeO2Loss}
  \includegraphics[height=0.8\columnwidth, angle=-90]{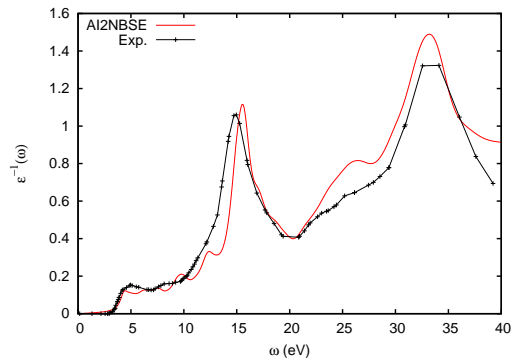}\vspace{3 mm}
  \caption{(Color online) Theoretical loss function of CeO$_2$ compared with experimental results.\cite{CeO2Loss} 
    Note that the first major peak at $\approx 15$ eV corresponds very
    well with the high binding energy satellite in the XPS, whereas
    the low energy satellite has no corresponding peak in the loss
    function.}
\end{figure}
\subsection{Extrinsic Losses}

Our treatment of extrinsic losses in the cumulant $C_k(t)$ is based on
the GW approximation to the cumulant, as discussed by Hedin and others
for the one-particle Green's function,
and leads to a Landau form similar to that for
intrinsic losses.  The difference is that the kernel is given by 
\begin{equation}
\beta_k(\omega) = \frac{1}{\pi} |{\rm Im}\, \Sigma_k(\omega+ \epsilon_k)|,
\end{equation}
where $\Sigma_k(\omega)$ is the electron self-energy calculated in the GW
approximation.\cite{KasRC,sky}
Formally the GW self-energy 
can be expressed in terms of fluctuation potentials as
$\Sigma (\omega) = \Sigma_q V^q \tilde g(\omega-\omega_q) V^q$.
In contrast to the CG, the spectral function from the GW approximation
only contains single boson excitations.  In this work we use the efficient
many-pole model self-energy\cite{MPSE} to calculate the kernel. The model
is based  on a representation of the dielectric function in
the screened coulomb interaction $W= \epsilon^{-1} v$
in terms of plasmon-like excitations,
and matches to a zero momentum-transfer loss function $L(\omega)$,
\begin{equation}
L(\omega) = -{\rm Im}\, \epsilon^{-1}(\omega).
\end{equation}
In this work $ L(\omega)$ is calculated via the first-principles
code AI2NBSE, which solves the 
GW/Bethe-Salpeter equation for valence excitations.\cite{AI2NBSE}
As an example, we show the calculated loss function for CeO$_2$
compared to experiment in Fig.\ 3.
The GW excitation spectrum in the cumulant from ${\rm Im}\, \Sigma_k$ is then
given by the MPSE model as a sum of plasmon pole self-energies, 
\begin{equation}
  {\rm Im}\, \Sigma_{k}(\omega) = \sum_{i} g_{i}\, {\rm Im}\,\Sigma^{i}_{k}(\omega;\omega_i),
\end{equation}
where $\Sigma^{i}(\omega, \omega_{i})$ is the 
the plasmon pole self-energy evaluated with plasmon frequency
$\omega_{i}$, and $g_{i}$ is the associated weight from the pole
representation of the loss function. 

\subsection{Interference Terms}


The calculation of the interference terms poses a number of technical
difficulties, especially in the treatment of recoil.
Although various approximations have been introduced,
including plasmon-pole models and semi-classical approximations,
\cite{hedin99review,bard85,Campbell2002,sokcevik79}
none is as yet fully satisfactory, and can lead to negative spectral
weight in some cases. On the other hand, one expects physically that
the kernel $\beta_K(\omega)$, like $\beta_c$ and $\beta_k$, should be
positive definite. Another complication with the interference terms is that the exchange of a boson between the photoelectron and hole should be associated with a change in photoelectron state, i.e., $k \rightarrow k+q$. Only if this is neglected (at least partially) can the particle-hole Green's function be considered diagonal, so that Eq.\ ({\ref{phxas}) is strictly valid (see Appendix A).
This approximation is not expected to capture all excitonic effects well, which is why we use a starting point that includes the static interaction and screening of the core-hole, and apply the convolution to incorporate non-adiabatic effects. In terms of fluctuation potentials the interference
contribution is given by the cross-terms in Eq.\ (\ref{perfectsquare})
\begin{equation}
\gamma_{ck}(\omega) = -2 \sum_{\bf q} \frac{V^{\bf q}_{\bf k k+q}
V^{{\bf q}*}_{cc}}{\omega_{\bf q} \omega_{\bf kq}}
\delta(\omega-\omega_{\bf kq}).
\end{equation}
Here
$\omega_{\bf kq}= \omega_{\bf q}+\epsilon_{\bf k-q}-\epsilon_{\bf k}$
 is the excitation energy
including recoil. Various approximations can be used in practical calculations. 
For example, the recoil effects can be approximated by neglecting the cross
terms in $\omega_{qk}$,  i.e., averaging over all directions $\hat {\bf q}$,
so that $\omega_{qk} \approx \omega_q +(1/2)q^2$.\cite{hedinrecoil80} We find
that for the plasmon pole model, neglecting recoil altogether is a reasonable
approximation for $k$ near $k_{F}$, although some of the integrals become
ill defined at large $k$ when recoil is neglected.
A similar recoil approximaion was used in deriving the cumulant expansion for
the one-particle Green's function.\cite{hedinrecoil80}
We find that for the plasmon pole model, this approximation is reasonable
for $k$ near $k_{F}$, although some of the integrals may become ill
defined at large $k$ when recoil is neglected.  As an alternative,
both the positive-definitiveness of the total kernel, and qualitative behavior
of the interference terms can be enforced by approximating the interference
term as
\begin{equation}
  \label{eq:inter}
  \beta_{ck}(\omega) = -2\lambda \sqrt{\beta_c(\omega)\beta_k(\omega)}.
\end{equation}
where $\lambda$ is an adjustable parameter.
Here we compare only calculations with $\lambda = 1$ or $\lambda=0$.
This {\it ad hoc} interpolation model ignores the phases in the interference
amplitudes  but preserves the correct limiting behavior:
$\lambda = 1$ gives the maximum possible interference, while
interference is neglected when $\lambda=0$ and vanishes
if either $\beta_c(\omega)$ or $\beta_k(\omega)$ is zero.  
We have also verified that this form gives a good approximation to the
satellite weight when compared to the interference term within the
plasmon pole approximation [Eq.\ (\ref{eq:ppinterf}) in Appendix
A].

In order to assess the quality of the above approximation, we calculated the
interference within the plasmon pole model near the Fermi momentum, which is
of particular importance for XANES. First, we used the interpolation model
above, and second, an approximation which neglects recoil (see Appendix A).
Fig.\ 4
shows that these two approximations are nearly identical near $k_F$.
\begin{figure} [ht]
\label{fig:inter}
\includegraphics[width=0.60\columnwidth, angle=-90]{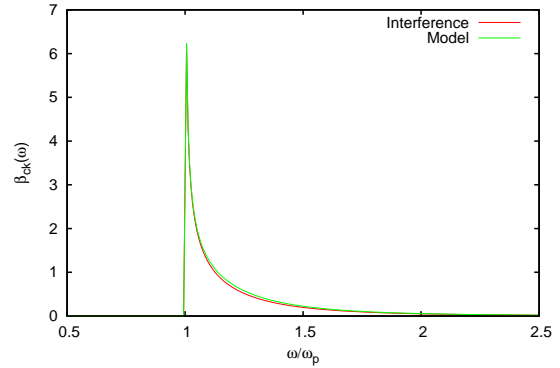}\vspace{3 mm}
\caption{(color online)
Comparison of calculated interference term $\beta_{ck}(\omega)$ (red) with
the interpolation model given by
Eq.~(\ref{eq:inter}) (green) for $k = k_{F}$. Both were calculated 
using the plasmon pole approximation and the recoil
approximation detailed in Appendix A.} 
\end{figure}


\begin{figure} [h]
\label{fig:xas-tm}
\includegraphics[width=0.60\columnwidth, angle=-90]{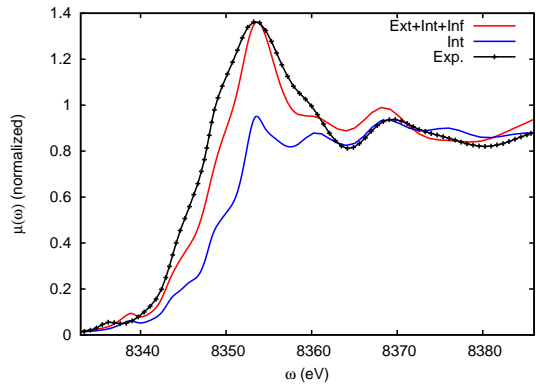}\vspace{3 mm}
\caption{(color online)
Ni K-edge XAS spectrum for NiO calculated with
intrinsic losses alone (blue), and all losses (red), compared to experiment (crosses).\cite{calandra2012}}
\end{figure}

The importance of interference effects in XAS is demonstrated in
Fig.\ 5, which shows a comparison between experiment and calculations of
the XAS of NiO with only intrinsic losses (blue) and with all losses (red).
Clearly the results agree remarkably well with experiment. They also
compare well with previous work,\cite{calandra2012,klevak2014}
although in those approaches the experimental XPS was used to
approximate the spectral function, and interference effects were
assumed to cancel the high energy plasmon entirely.
For this example, the amplitude of the core-hole potential was adjusted
in order to achieve reasonable agreement with satellite intensity in the XPS.

\section{X-ray Spectra}
Finally in this section we present llustrative results of our approach for
the XPS and XAS 
for a variety of materials. The single particle XAS spectra were calculated
using the FEFF9 code,\cite{feffcr} which was then
convolved with the spectral function derived above.

\begin{figure} [ht]
\label{fig:xas-al}
\includegraphics[width=0.60\columnwidth, angle=-90]{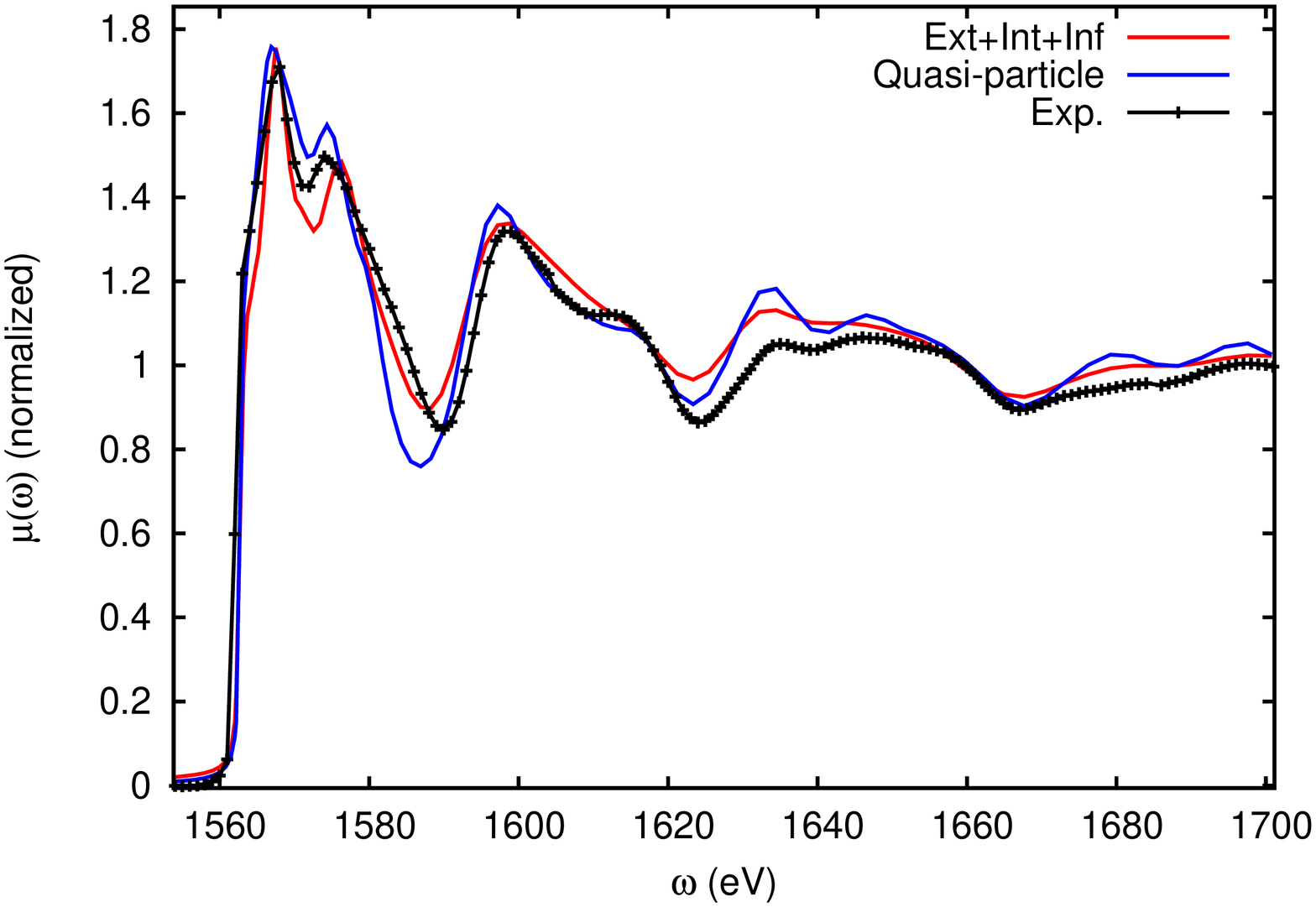}\vspace{3 mm}
\includegraphics[height=0.8\columnwidth, angle=-90]{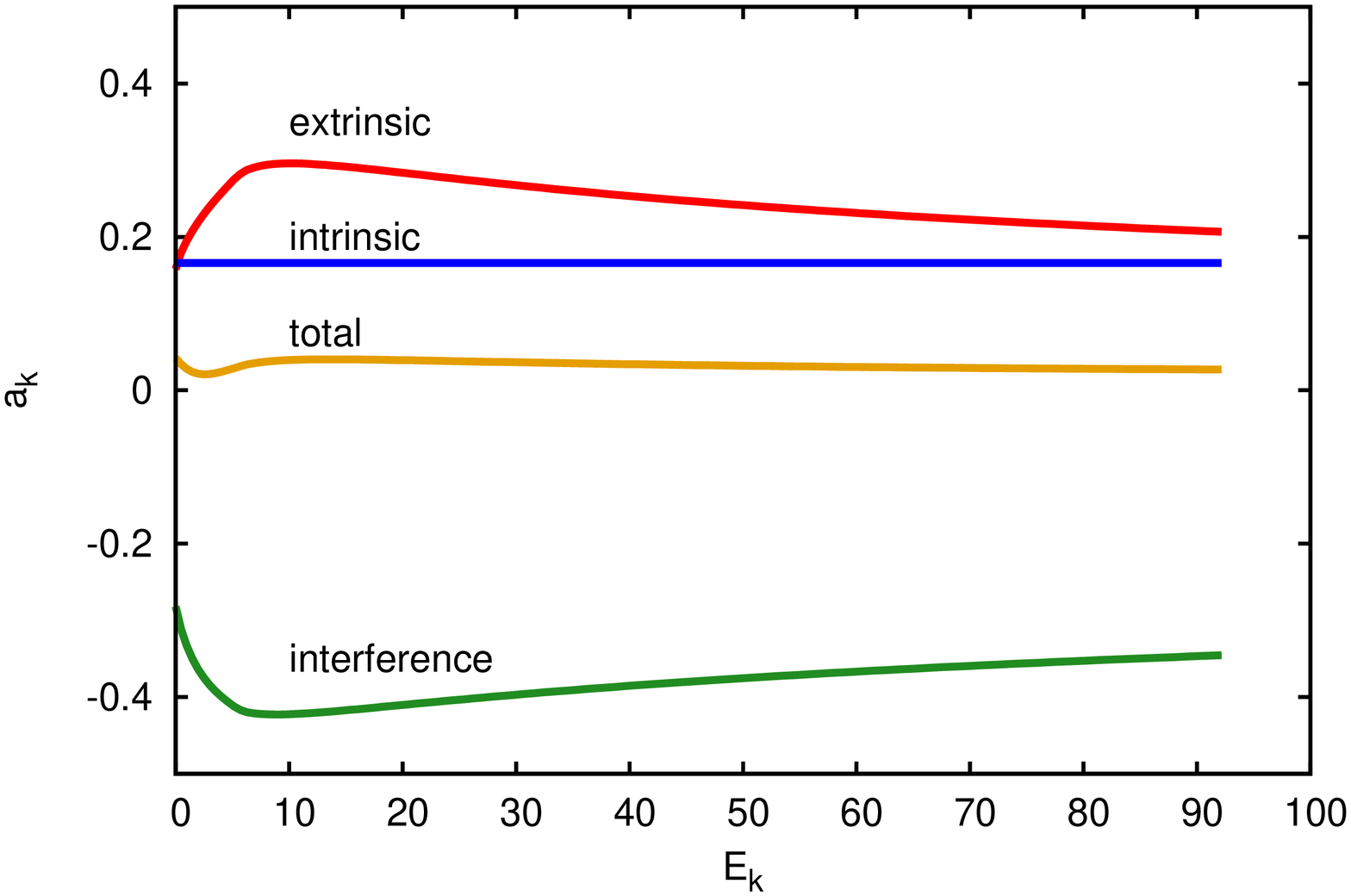}\vspace{3 mm}
\caption{(Color online)
(top) Theoretical Al K-edge XAS spectrum compared to the quasi-particle theory in
this work and
experimental data;\cite{lagarde99} and
(bottom) total satellite weight $a_{k}$ for Al
and different contributions as a function of the particle energy $E_{k}$.
Note that the total satellite strength has the expected structure, being
small at threshold and exhibiting significant cancellation among the
various contributions.
}
\end{figure}

\subsection{K-edge Al}
As a first example, we show the experimental XAS for 
fcc Al metal compared to the calculated results including
the cumulant convolution, and those of the single particle
calculation (Fig.\ 6 top). Both calculations both agree fairly well
with experiment, although the single particle spectrum
does not contain enough broadening at about 1590
eV, where the dip is too large. 
The figure (bottom panel) also shows the separate contributions from the
intrinsic, extrinsic, and interference satellites.
Note that the interference terms nearly completely cancel
the weight of the extrinsic and intrinsic satellites,
and return that weight to the quasiparticle peak, showing that in this
case, the adiabatic approximation is reasonable.


\subsection{Elemental transition metals}

\begin{figure}[ht]
  \label{fig:metals_XPS}
  \includegraphics[height=0.8\columnwidth, angle=-90]{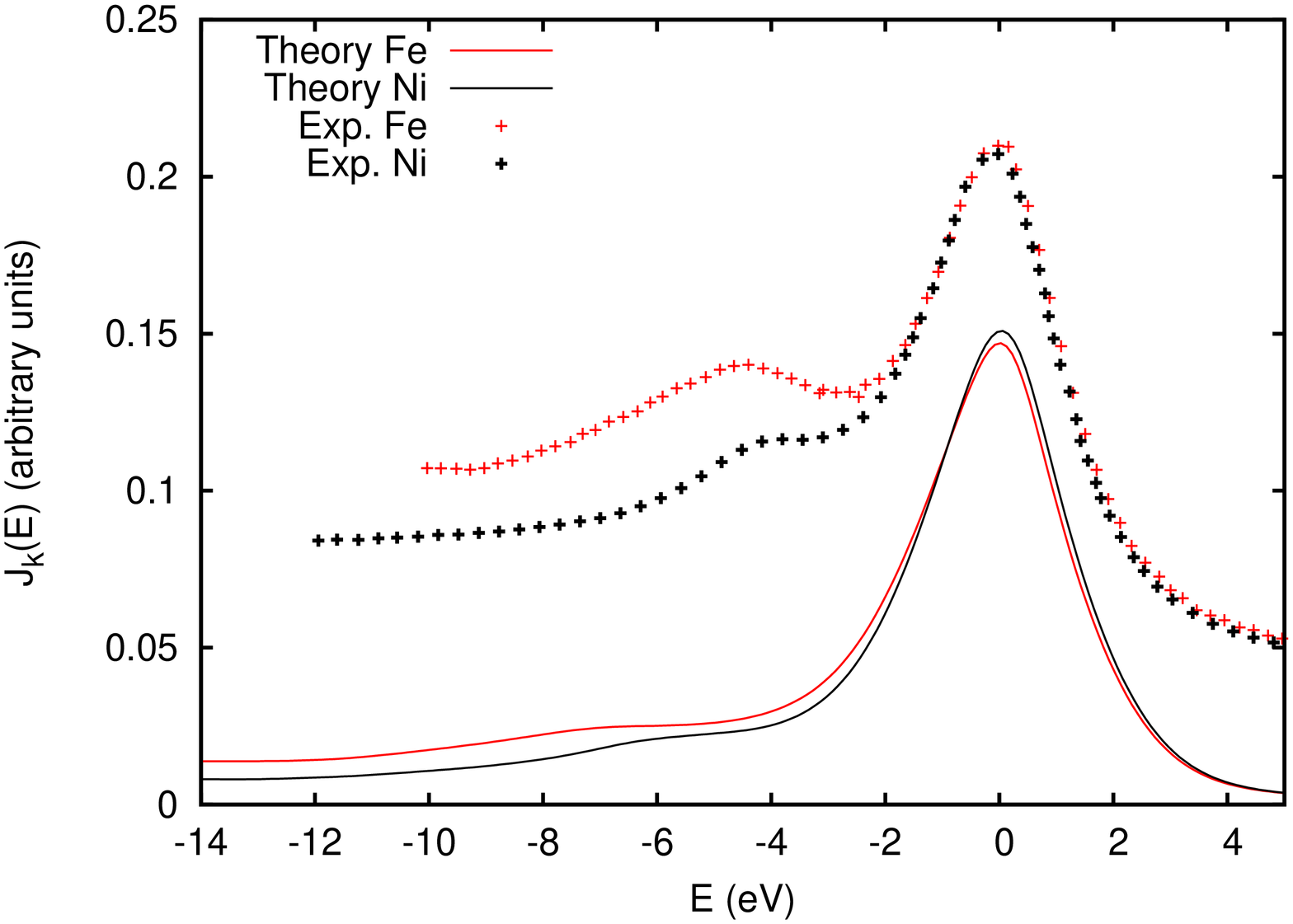}
\includegraphics[width=0.8\columnwidth]{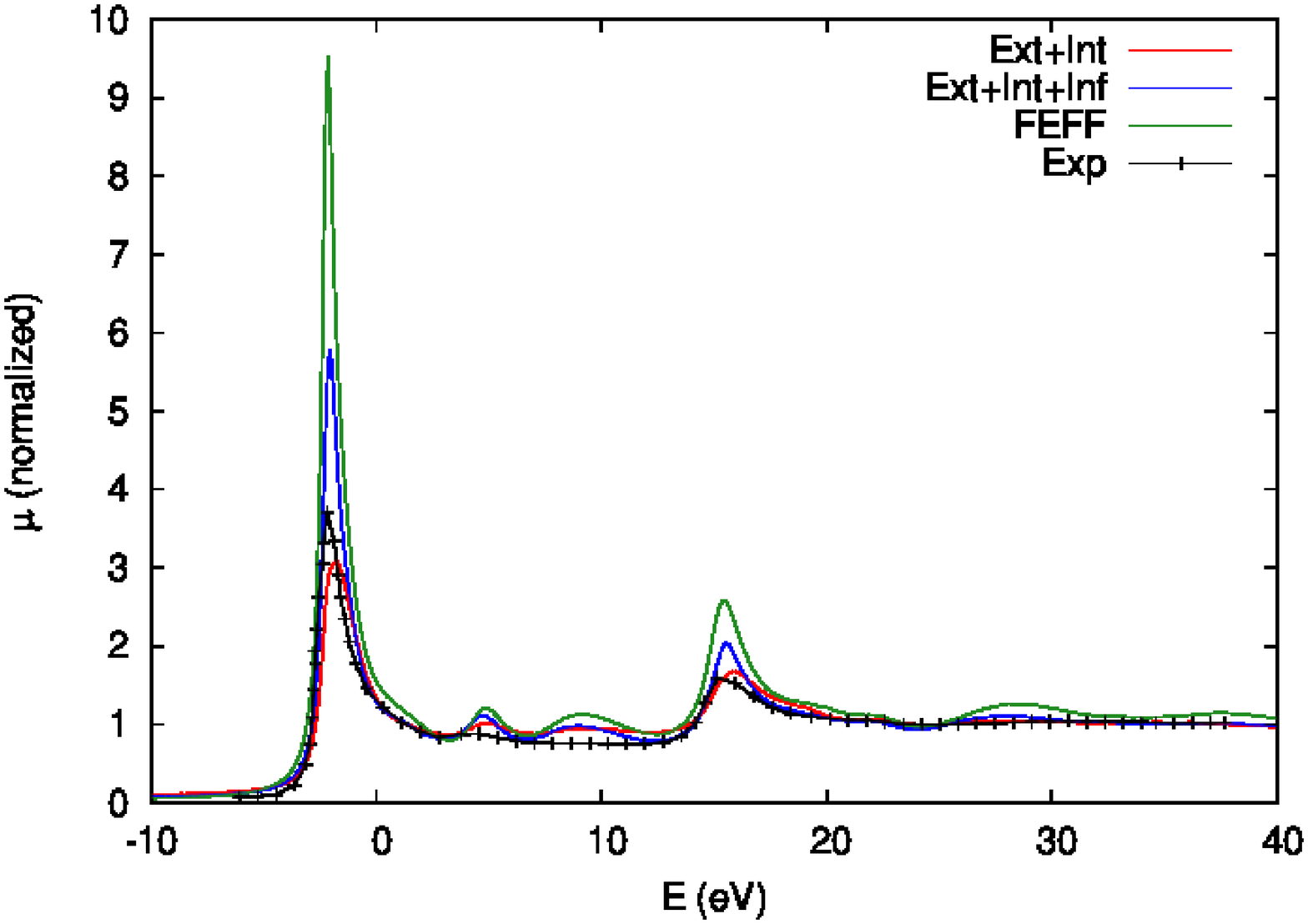}
\caption{(Color online) Calculated spectral function of Fe, and Ni
compared to experimental Ni 1s and Fe 3s
XPS data (top).\cite{karis2008,xu1995} The calculated Ni K-edge XANES
is also compared to experiment (bottom).\cite{chen1991}
}
\end{figure}

As an application of the theory to $d$-electron materials, we 
present results for the core-hole spectral function and photocurrent 
of elemental transition metals Fe and Ni from Eq.\ (\ref{photocurrent})
in Fig.\ 7.
For these systems a 6 eV satellite observed in some experimental spectra
has been of great interest, but it's origin has been controversial,
especially for Ni.  Interestingly the RT-TDDFT approach also
yields satellites around 6 eV; however, their amplitude is smaller
than that observed in experiment. In addition, the experimental XPS data for Ni shows a satellite closer to 4 eV. This energy/amplitude mismatch is
also apparent in the comparison of theory and experiment in Fe. However, the qualitative differences between theory and experiment are consistent,
where the satellite moves toward the main peak with decreased amplitude
going from Fe to Ni.

Next we use the particle-hole spectral function  to
calculate the XAS using the convolution in Eq.~(\ref{phxas}).
As Fig.~7 indicates, the effect of the satellites
generally adds to the asymmetry of the
edge peak, leading better agreement with experiment than
the single particle spectrum. In this case, ignoring interference
effects ($\lambda = 0$) seems to be a reasonable approximation,
although this agreement may be due in part to
the neglect of multiplet effects that mix the edges.

\subsection{Charge-transfer satellites }

\begin{figure}[ht]
  \label{fig:ceo2_XS}
  \includegraphics[height=0.8\columnwidth, angle=-90]{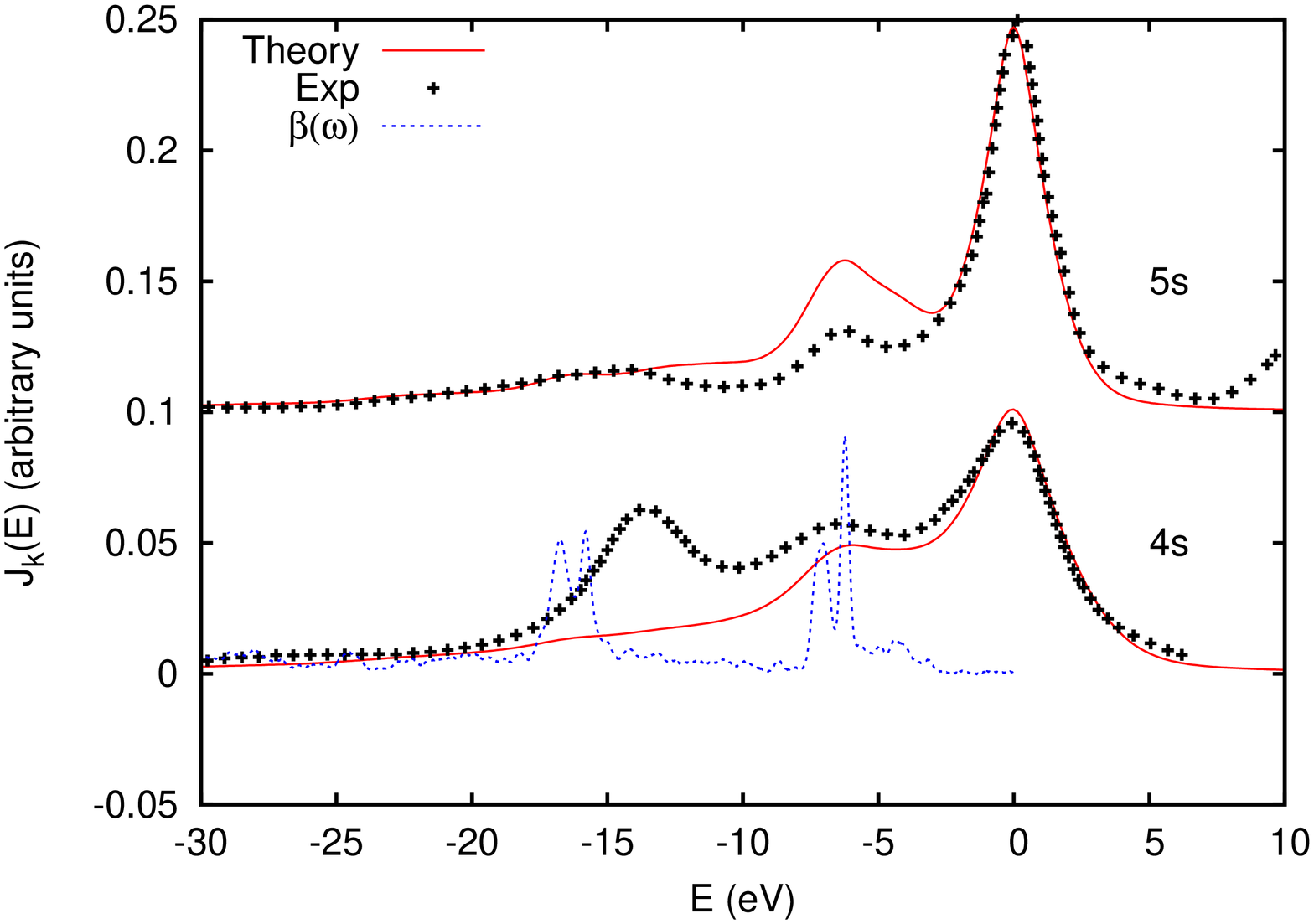}
\includegraphics[height=0.8\columnwidth, angle=-90]{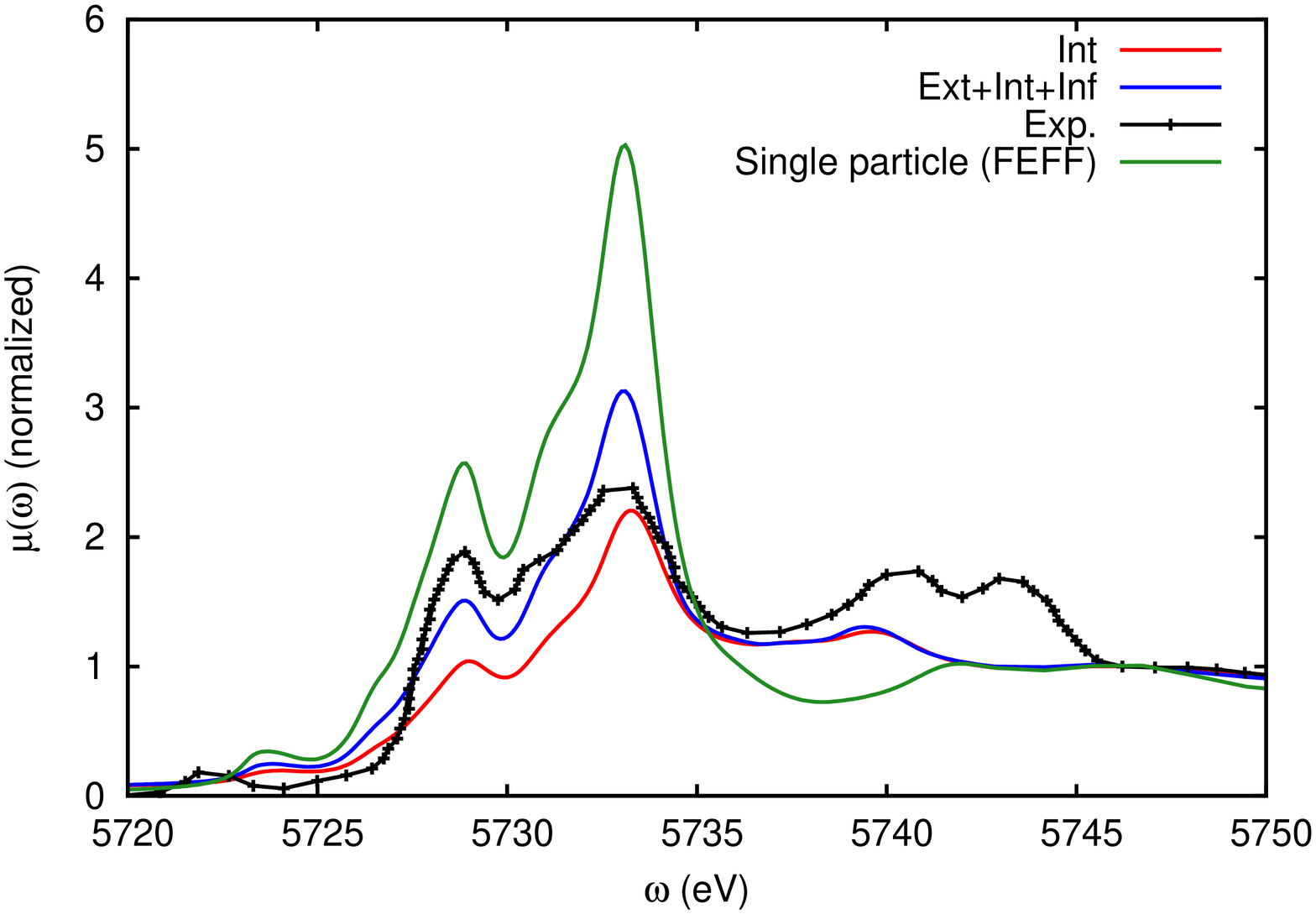}
  \caption{(Color online)
Theoretical Ce 4s and 5s XPS of CeO$_2$ compared with
experiment\cite{CeO2XPS} (top) and similarly the Ce L$_3$ XAS spectrum
(bottom) at various levels of approximation,
 also compared to experiment.\cite{kotani2012} 
}
\end{figure}

Charge-transfer (CT) excitations are particularly strong in transition
metal oxides, and have been the subject of numerous
investigations.\cite{lgh,calandra2012,KRC}
Here we discuss the application of our particle-hole cumulant approach
to CT satellites, namely the $f$-electron system CeO$_2$, again
following the treatment of the intrinsic losses in XPS by KVRC.
The top plot in Fig.\ 8
shows the core-hole spectral function
compared to the experimental Ce 4$s$ and 5$s$ XPS of CeO$_2$. The
agreement in peak position is reasonable, although
the qualitative change in satellite amplitudes in going from the 4s to 5s
hole is not reproduced.
There could be several reasons for this discrepancy:
first, our present calculations ignore the shape of the core-state, and
thus the core-hole potential may not be accurate; and  second --and possibly
more important for the case of CeO$_2$ -- the core-hole potential is
assumed to be a static coulomb potential and exchange is ignored.
Third, the 5s states are
treated as core states in the calculation, and should probably be
promoted into the valence for this system. Finally, frustrated Auger
configurations can also play a role in the spectrum, but are not treated
here.\cite{bagus2010} 
Fig.\ 8 (bottom)
shows our calculations of the XAS 
of CeO$_2$ calculated at
various levels of theory and compared with experiment.
Interestingly the agreement with experiment is quite reasonable with
only intrinsic losses, suggesting that extrinsic losses and
interference have a minor effect on the XAS.
Consequently these CT systems represent cases
where the cancellation of the intrinsic and extrinsic losses is generally
incomplete at threshold, and that intrinsic losses are essential for a 
quantitative treatment.
This is supported by investigations of CT systems using the 3-state
model of Lee, Hedin, and
Gunnarsson,\cite{lgh,klevak2014,calandra2012} where the sudden
limit is reached at relatively low energies, and that the interference
is in fact constructive near threshold. This constructive interference
is missing in our model of interference (at least with $\lambda = 1$),
and may be one reason for some missing amplitude in our
CT excitation in CeO$_2$.

\begin{figure}[ht]
  \includegraphics[height=0.8\columnwidth, angle=-90]{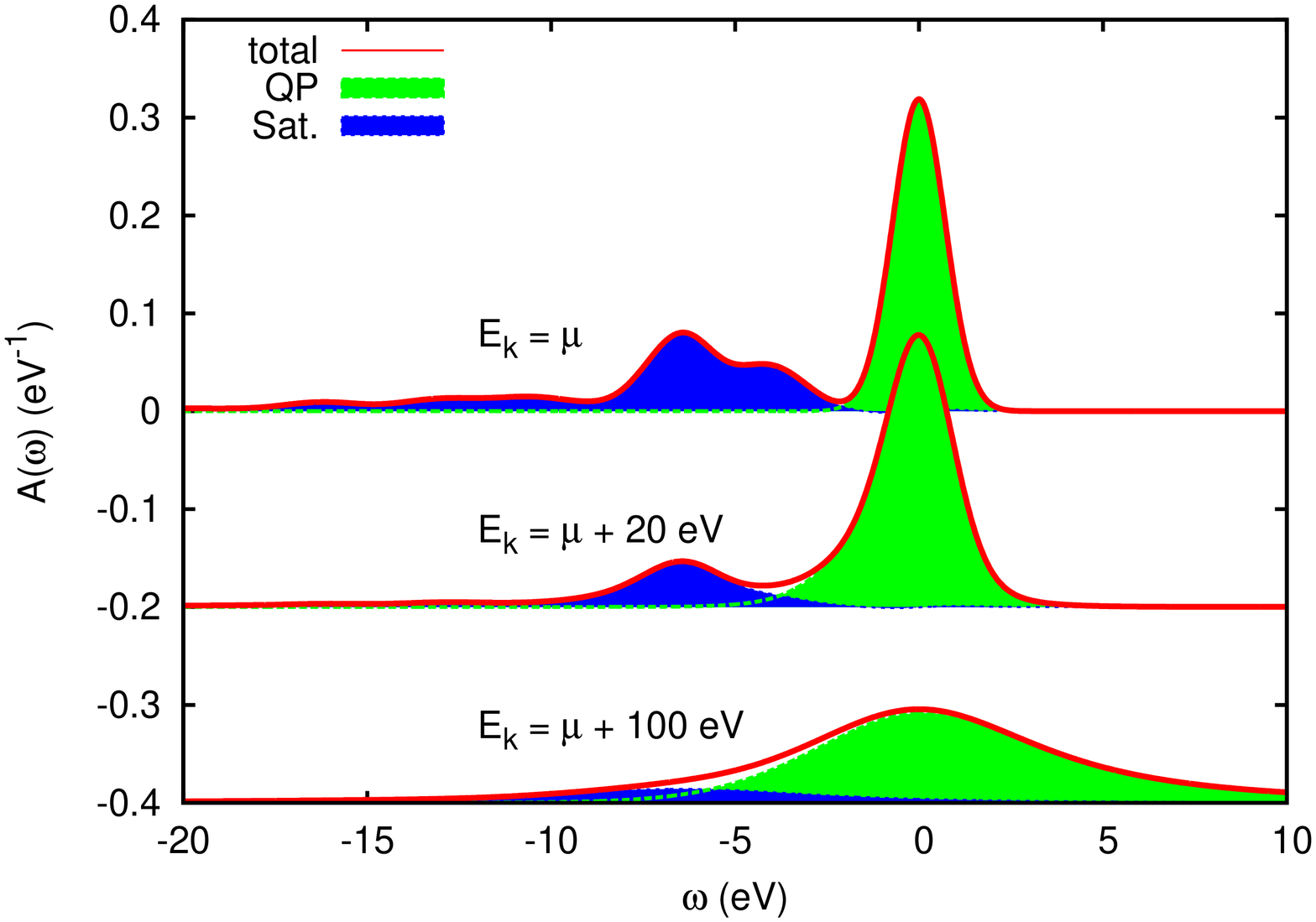}\vspace{1 mm}
  \includegraphics[height=0.8\columnwidth, angle=-90]{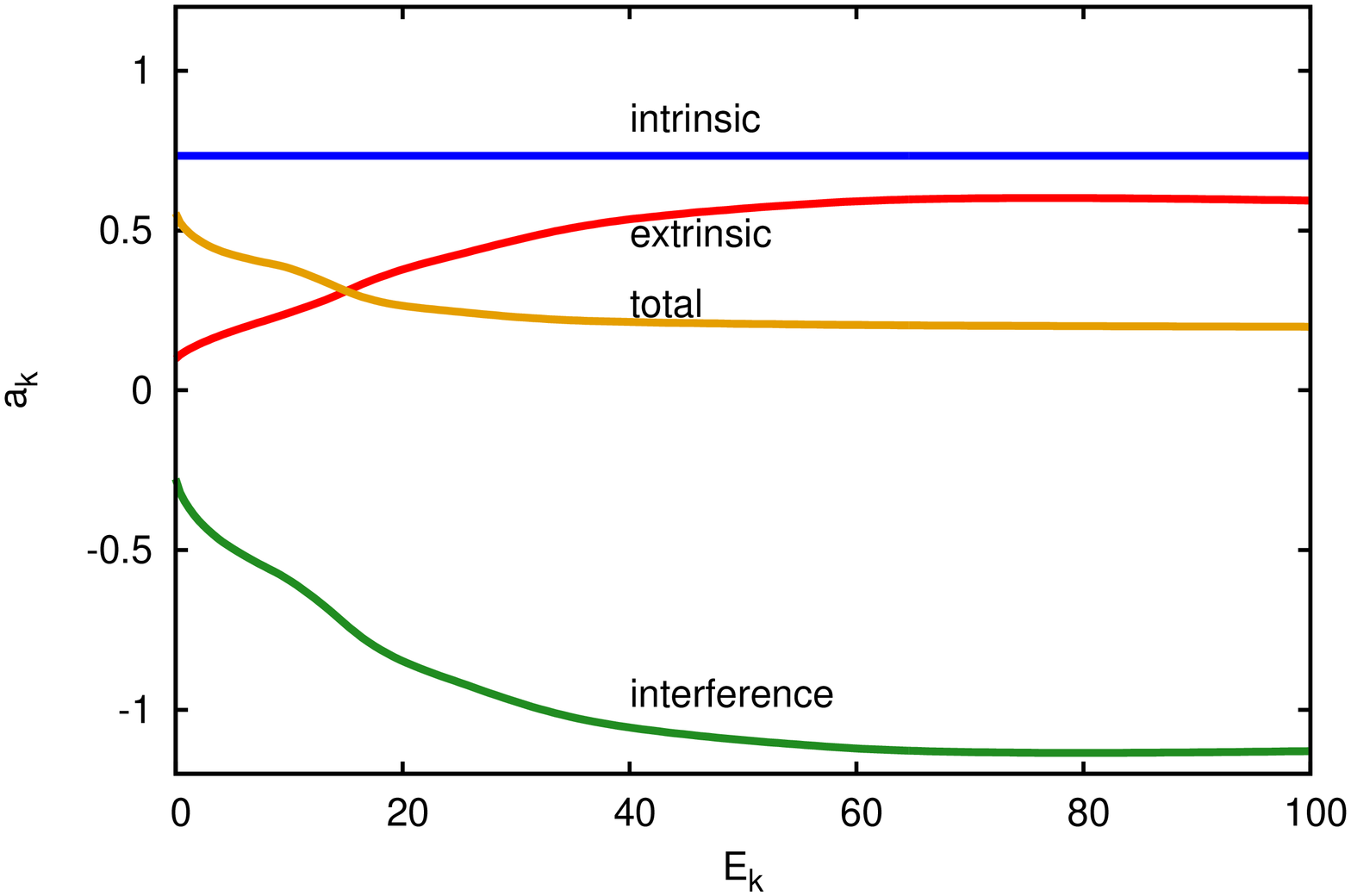}
  \caption{(Color online)
 For reference the
particle-hole spectral function $\tilde A_K(\omega)$ for CeO$_2$ (top) and
the satellite weights of the individual contributions (bottom) are also
given.}
\end{figure}

\section{Conclusions}
\label{sect_conclusions}
We have developed both the theory and a practical approach for calculating
inelastic losses due to
intrinsic, extrinsic and interference effects in x-ray spectra within
a generalized particle-hole cumulant expansion, and a partition of
the cumulant into extrinsic, intrinsic and interference contributions. 
These losses are
included in the spectra in terms of a convolution with a particle-hole
spectral function that accounts for their energy dependence.
The cumulant approach simplifies the formalism and facilitates practical
calculations.  Here the core-hole cumulant is calculated via a
real-time real-space approach, while an efficient many-pole model
self-energy is used to obtain the extrinsic part. Interference effects
are approximated using an interpolation model. This model avoids
the complications of recoil effect which can lead to a number of technical
problems such as unphysical negative spectral weight at some energies.
The
approach is complementary to, and can be used to correct various methods
for calculating XAS and XPS that ignore inelastic losses and satellites
e.g., in a post-processing step.
The cumulant theory also elucidates both their behavior and the differences
between the spectral functions for XAS and XPS which may be important
to their interpretation. 
In contrast to the relatively sharp features of independent particle
spectral function for the ground state, the particle-hole spectral function
exhibits significant breadth and asymmetry of both the quasi-particle peak
and the satellite contributions well above the Fermi energy.
The cumulant approach can also account for edge singularities in the spectrum.
Physically, the treatment of inelastic losses here is analogous to an
``excitonic polaron," i.e., the interaction of the particle-hole created
in photoexcitation with the density fluctuations produced by the 
particle-hole system.  This is in contrast to the ``electronic polaron"
described by the GW approximation,\cite{hedin99review} where the
single-particle excitations arise from the generally stronger
density fluctuations  due to to a core-hole or a photoelectron.
 For the L-edge XAS of Fe and Ni, the spectral function reduces the whiteline
significantly and increases the asymmetry of the peak, bringing the
theoretical curve closer to experiment, and giving good
agreement, especially in Ni. However, there is missing weight at
roughly 5-7 eV, indicating that the strength of the satellites in this
region is underestimated with the current approximation.
As in CHRB, we find that interference effects play
a large role in reducing the effects of many-body excitations, especially
in XAS. However, in general, we find the  cancellation to be incomplete,
especially in charge-transfer systems, where the intrinsic satellites
are dominant. For these cases the inclusion of both interference terms
and extrinsic losses are usually
essential to provide a quantitative treatment of
the near edge spectrum. Finally, although
effects of thermal vibrations are not discussed in detail here,
they can be folded 
into the calculation of the single particle spectrum $\mu^0(\omega)$.
These effects can be approximated via the use of Debye-Waller factors as
in the 
real-space multiple scattering code FEFF9,\cite{feffcr,dmdw} by
convolving with an effective quasiparticle spectral
function,\cite{Shirleyetal16} or by performing a
configurational average over MD snapshots. The effects of phonons on
the satellite structure is expected to be rather weak, and is in many
cases obscured by the large core-hole lifetime, but can also be
treated via cumulant
methods.\cite{story2014,gunnarsson1994}
 
\noindent Acknowledgments - We thank G. Bertsch, C. Draxl, C. Fadley,
A. Lee, L. Reining, E. Shirley, and T. Devereaux for comments and suggestions.
This work was supported by DOE Grant DE-FG03-97ER45623 (JJR and JJK);
One of us (JC) was supported by the 2015 NSF REU summer program at UW.  
  
\appendix
\section{Inelastic losses XAS}

Here we briefly summarize the derivation of the generalized particle-hole
cumulant. We start with the treatment in CHRB\cite{Campbell2002} where
the many-body XAS is given by
\begin{equation}
  \mu(\omega) = -\frac{1}{\pi}{\rm Im}\langle c|d^{\dagger}P\, g_{\rm eff}(\omega+E_{c})\, P d |c\rangle.
\end{equation}
effective Green's function $g_{\rm eff}$ can be conveniently expressed in
terms of the fluctuation potentials $V^{\bf q}$ which diagonalize the dielectric
response,
\begin{equation}
   {\rm Im}\, W({\bf r},{\bf r'},\omega)  = \sum_{\bf q} V^{\bf q}({\bf r})
V^{*{\bf q}}({\bf r'}) \delta(\omega-\omega_{\bf q}).
\end{equation}
The effective Green's function then becomes
\begin{align}
  g_{\rm eff}(\omega) &= e^{-a}\left[\tilde g(\omega) + \sum_{\bf q}\left|\frac{V^{\bf q}_{cc}}{\omega_{\bf q}}\right|^{2}\tilde g(\omega-\omega_{\bf q})\right. \nonumber \\
  &\left.-2\sum_{\bf q}\frac{V^{\bf q}_{cc}}{\omega_{\bf q}}\tilde g(\omega-\omega_{\bf q})V^{\bf q}\tilde g(\omega)\right],
\end{align}
where $\tilde g(\omega)$ is the photoelectron Green's function
in the presence of the core-hole including extrinsic interactions and
$a=\sum_{\bf q} |V_{cc}^q/\omega_q|^2$.
expression simplifies considerably in the time domain. for example
the second term which characterizes the satellite contribution from
intrinsic losses as
$\sum_{\bf q} |V_{cc}^q/\omega_q|^2 \exp(i\omega_q t).$ Similarly
 if we now express $g_{\rm eff}$ to second order in the fluctuation
potentials which are implicit in the definition of $g(\omega)$, we obtain
\begin{eqnarray}
 g_{\rm eff}(\omega) =   g^{0}(\omega) &+& \left[
  \sum_{\bf q} g^{0}(\omega)V^{\bf q}g^{0}(\omega-\omega_{\bf q})
                         V^{\bf q}g^{0}(\omega)  \right. \nonumber \\
  &+& \sum_{\bf q}\left|\frac{V^{\bf q}_{cc}}{\omega_{\bf q}} \right|^{2}
     g^{0}(\omega-\omega_{\bf q}) \nonumber \\
 &-& \left. 2\sum_{\bf q}\frac{V^{\bf q}_{cc}}{\omega_{\bf q}}
     g^{0}(\omega-\omega_{\bf q})V^{\bf q}g^{0}(\omega)\right].
\end{eqnarray}
Now we can follow the procedure of Gunnarson to derive a second order cumulant approximation for the effective Green's function in the time domain  which has
the structure of a particle-hole Green's function,
\begin{eqnarray}
G_K(t) &=&  g^0_c(t) g^0_k(t) e^{C_K(t)} \\
    C_K(t) &=& \int d\omega \gamma_K(\omega)
      \left(e^{i \omega t} - i \omega t - 1\right) \nonumber \\
\gamma_K(\omega) &=& \gamma_k(\omega) + \gamma_c(\omega) + \gamma_{kc}(\omega).
\end{eqnarray}
The kernels $\gamma_c$ and $\gamma_k$ of the intrinsic and extrinsic
cumulants are given by
\begin{align}
\gamma_c(\omega) &= \sum_{\bf q}\frac{|V^{\bf q}_{cc}|^{2}}{\omega_{\bf q}^2}
      \delta(\omega-\omega_{\bf q}), \nonumber \\
\gamma_k(\omega) &= \sum_{\bf q} \frac{|V^{\bf q}_{\bf k k+q}|^{2}}{\omega_{\bf kq}^{2}}\delta(\omega - \omega_{\bf kq})  \nonumber \\
&= \frac{|{\rm Im}\Sigma_{\bf k}(\omega+E_{\bf k})|}{\pi \omega^{2}}.
\end{align}
The inteference term can be derived similarly, although a further approximation of constant matrix elements must be made, namely
$\langle k | d | c \rangle = \langle k + q | d | c \rangle$. This yields
the interference kernel
\begin{equation}
  \gamma_{kc}(\omega) = -2 \sum_{\bf q} \frac{V^{\bf q}_{cc}
V^{\bf q}_{\bf k k+q}}{\omega\, \omega_{\bf kq}} \delta(\omega-\omega_{\bf kq}),
\end{equation}
where $\omega_{\bf kq} = \omega_{\bf q} + E_{\bf k+q}-E_{\bf k}$.
%

As a concrete example, we illustrate the result with the plasmon pole
approximation for the dielectric function. In that case the
fluctuation potentials are plane waves, $V^{\bf q}({\bf r}) = 
V^{q}_{0} \exp({\bf k}\cdot{\bf r})$. Thus we have $V^{\bf q}_{cc} = 
V^{q}_{0}$ and $V^{\bf q}_{kk'} = v^{q}_{0}\delta_{\bf k',k+q}$. 
The integrals over ${\bf q}$ can be performed, to find explicit
expressions for the contributions to $\gamma_K(\omega)$, however,
we find that this solution is
plagued by unphysical behavior such as negative spectral density. If
we instead neglect recoil ($\epsilon_{\bf k+q} \approx \epsilon_{\bf
k})$ in the delta functions, we can then write the
complete excitation spectrum as a perfect square, which will ensure a
positive definite result. In that case we find
\begin{eqnarray}
\gamma_{k}(\omega) &=& \frac{
\omega_{p}^2\theta(\omega-\omega_p)}{ \pi \omega
\sqrt{2(\omega-\omega_p)}} \nonumber \\
&\times&\frac{1}{\left[(2\omega-\omega_p)^2-2k^2(\omega-\omega_p)\right]},
\nonumber \\ 
\gamma_c(\omega) &=&
\frac{\omega_{p}^{2}\theta(\omega-\omega_{p})}{\pi
\sqrt{2(\omega-\omega_{p})}\omega^{3}}, \nonumber \\
\gamma_{kc}(\omega) &=& \frac{\omega_{p}^2\theta(\omega-\omega_p)}{4\pi k \omega^2
(\omega-\omega_p)} \nonumber \\
&\times&\ln \left[\frac{2\omega - \omega_p + k\sqrt{2(\omega-\omega_p)}}{2\omega - \omega_p - k\sqrt{2(\omega-\omega_p)}}\right],
\label{eq:ppinterf}
\end{eqnarray}
where for simplicity we have taken the plasmon dispersion to be
$\omega_{q}=\omega_{p}+1/2 q^{2}$ and
$V_0^q = [v_q\omega_p^2/2\omega_q)]^{1/2}$.
Unfortunately, these expressions are only valid for low photoelectron
momentum $k$, 
since the integrals are ill defined for large $k$ when recoil is
neglected. However, 
we use them only to evaluate our approximation
for the interference term in reference to XANES calculations, where the
photoelectron momentum is relatively low.


\vspace{2 in}


\begin{thebibliography}{51}
\expandafter\ifx\csname natexlab\endcsname\relax\def\natexlab#1{#1}\fi
\expandafter\ifx\csname bibnamefont\endcsname\relax
  \def\bibnamefont#1{#1}\fi
\expandafter\ifx\csname bibfnamefont\endcsname\relax
  \def\bibfnamefont#1{#1}\fi
\expandafter\ifx\csname citenamefont\endcsname\relax
  \def\citenamefont#1{#1}\fi
\expandafter\ifx\csname url\endcsname\relax
  \def\url#1{\texttt{#1}}\fi
\expandafter\ifx\csname urlprefix\endcsname\relax\def\urlprefix{URL }\fi
\providecommand{\bibinfo}[2]{#2}
\providecommand{\eprint}[2][]{\url{#2}}

\bibitem[{\citenamefont{Hedin}(1965)}]{Hedin:1}
\bibinfo{author}{\bibfnamefont{L.}~\bibnamefont{Hedin}},
  \bibinfo{journal}{Phys. Rev.} \textbf{\bibinfo{volume}{139}}
  (\bibinfo{year}{1965}).

\bibitem[{\citenamefont{Kas et~al.}(2007)\citenamefont{Kas, Sorini, Prange,
  Cambell, Soininen, and Rehr}}]{MPSE}
\bibinfo{author}{\bibfnamefont{J.~J.} \bibnamefont{Kas}},
  \bibinfo{author}{\bibfnamefont{A.~P.} \bibnamefont{Sorini}},
  \bibinfo{author}{\bibfnamefont{M.~P.} \bibnamefont{Prange}},
  \bibinfo{author}{\bibfnamefont{L.~W.} \bibnamefont{Cambell}},
  \bibinfo{author}{\bibfnamefont{J.~A.} \bibnamefont{Soininen}},
  \bibnamefont{and} \bibinfo{author}{\bibfnamefont{J.~J.} \bibnamefont{Rehr}},
  \bibinfo{journal}{Phys. Rev. B} \textbf{\bibinfo{volume}{76}},
  \bibinfo{pages}{195116} (\bibinfo{year}{2007}).

\bibitem[{\citenamefont{Hedin}(1999)}]{hedin99review}
\bibinfo{author}{\bibfnamefont{L.}~\bibnamefont{Hedin}}, \bibinfo{journal}{J.
  Phys.: Condens. Matter} \textbf{\bibinfo{volume}{11}}, \bibinfo{pages}{R489}
  (\bibinfo{year}{1999}).

\bibitem[{\citenamefont{Hedin et~al.}(1998)\citenamefont{Hedin, Michiels, and
  Inglesfield}}]{hmi}
\bibinfo{author}{\bibfnamefont{L.}~\bibnamefont{Hedin}},
  \bibinfo{author}{\bibfnamefont{J.}~\bibnamefont{Michiels}}, \bibnamefont{and}
  \bibinfo{author}{\bibfnamefont{J.}~\bibnamefont{Inglesfield}},
  \bibinfo{journal}{Phys Rev.B} \textbf{\bibinfo{volume}{58}},
  \bibinfo{pages}{15565} (\bibinfo{year}{1998}).

\bibitem[{\citenamefont{Campbell et~al.}(2002)\citenamefont{Campbell, Hedin,
  Rehr, and Bardyszewski}}]{Campbell2002}
\bibinfo{author}{\bibfnamefont{L.}~\bibnamefont{Campbell}},
  \bibinfo{author}{\bibfnamefont{L.}~\bibnamefont{Hedin}},
  \bibinfo{author}{\bibfnamefont{J.~J.} \bibnamefont{Rehr}}, \bibnamefont{and}
  \bibinfo{author}{\bibfnamefont{W.}~\bibnamefont{Bardyszewski}},
  \bibinfo{journal}{Phys. Rev. B} \textbf{\bibinfo{volume}{65}},
  \bibinfo{pages}{064107} (\bibinfo{year}{2002}).

\bibitem[{\citenamefont{Triguero et~al.}(1998)\citenamefont{Triguero,
  Pettersson, and \AA{}gren}}]{stobe}
\bibinfo{author}{\bibfnamefont{L.}~\bibnamefont{Triguero}},
  \bibinfo{author}{\bibfnamefont{L.}~\bibnamefont{Pettersson}},
  \bibnamefont{and}
  \bibinfo{author}{\bibfnamefont{H.}~\bibnamefont{\AA{}gren}},
  \bibinfo{journal}{J. Phys. Chem. A} \textbf{\bibinfo{volume}{102}},
  \bibinfo{pages}{10599} (\bibinfo{year}{1998}).

\bibitem[{\citenamefont{Vinson et~al.}(2011)\citenamefont{Vinson, Rehr, Kas,
  and Shirley}}]{ocean}
\bibinfo{author}{\bibfnamefont{J.}~\bibnamefont{Vinson}},
  \bibinfo{author}{\bibfnamefont{J.~J.} \bibnamefont{Rehr}},
  \bibinfo{author}{\bibfnamefont{J.~J.} \bibnamefont{Kas}}, \bibnamefont{and}
  \bibinfo{author}{\bibfnamefont{E.~L.} \bibnamefont{Shirley}},
  \bibinfo{journal}{Phys.~Rev.~B} \textbf{\bibinfo{volume}{83}},
  \bibinfo{pages}{115106} (\bibinfo{year}{2011}).

\bibitem[{\citenamefont{Puschnig and Ambrosch-Draxl}(2002)}]{exciting}
\bibinfo{author}{\bibfnamefont{P.}~\bibnamefont{Puschnig}} \bibnamefont{and}
  \bibinfo{author}{\bibfnamefont{C.}~\bibnamefont{Ambrosch-Draxl}},
  \bibinfo{journal}{Phys. Rev. B} \textbf{\bibinfo{volume}{66}},
  \bibinfo{pages}{165105} (\bibinfo{year}{2002}).

\bibitem[{\citenamefont{Rehr et~al.}(1978)\citenamefont{Rehr, Stern, Martin,
  and Davidson}}]{rehr78}
\bibinfo{author}{\bibfnamefont{J.~J.} \bibnamefont{Rehr}},
  \bibinfo{author}{\bibfnamefont{E.~A.} \bibnamefont{Stern}},
  \bibinfo{author}{\bibfnamefont{R.~L.} \bibnamefont{Martin}},
  \bibnamefont{and} \bibinfo{author}{\bibfnamefont{E.~R.}
  \bibnamefont{Davidson}}, \bibinfo{journal}{Phys.~Rev.~B}
  \textbf{\bibinfo{volume}{17}}, \bibinfo{pages}{560} (\bibinfo{year}{1978}).

\bibitem[{\citenamefont{Bagus et~al.}(2010{\natexlab{a}})\citenamefont{Bagus,
  Nelin, Ilton, Baron, Abbott, Primorac, Kuhlenbeck, Shaikhutdinov, and
  Freund}}]{bagus2010}
\bibinfo{author}{\bibfnamefont{P.}~\bibnamefont{Bagus}},
  \bibinfo{author}{\bibfnamefont{C.}~\bibnamefont{Nelin}},
  \bibinfo{author}{\bibfnamefont{E.}~\bibnamefont{Ilton}},
  \bibinfo{author}{\bibfnamefont{M.}~\bibnamefont{Baron}},
  \bibinfo{author}{\bibfnamefont{H.}~\bibnamefont{Abbott}},
  \bibinfo{author}{\bibfnamefont{E.}~\bibnamefont{Primorac}},
  \bibinfo{author}{\bibfnamefont{H.}~\bibnamefont{Kuhlenbeck}},
  \bibinfo{author}{\bibfnamefont{S.}~\bibnamefont{Shaikhutdinov}},
  \bibnamefont{and} \bibinfo{author}{\bibfnamefont{H.-J.}
  \bibnamefont{Freund}}, \bibinfo{journal}{Chem. Phys. Lett.}
  \textbf{\bibinfo{volume}{487}}, \bibinfo{pages}{237 }
  (\bibinfo{year}{2010}{\natexlab{a}}).

\bibitem[{\citenamefont{de~Groot and Kotani}(2008)}]{degrootbook}
\bibinfo{author}{\bibfnamefont{F.}~\bibnamefont{de~Groot}} \bibnamefont{and}
  \bibinfo{author}{\bibfnamefont{A.}~\bibnamefont{Kotani}},
  \emph{\bibinfo{title}{Core Level Spectroscopy of Solids}}
  (\bibinfo{publisher}{CRC Press}, \bibinfo{year}{2008}).

\bibitem[{\citenamefont{Casula et~al.}(2012)\citenamefont{Casula, Rubtsov, and
  Biermann}}]{biermann12}
\bibinfo{author}{\bibfnamefont{M.}~\bibnamefont{Casula}},
  \bibinfo{author}{\bibfnamefont{A.}~\bibnamefont{Rubtsov}}, \bibnamefont{and}
  \bibinfo{author}{\bibfnamefont{S.}~\bibnamefont{Biermann}},
  \bibinfo{journal}{Phys. Rev. B} \textbf{\bibinfo{volume}{85}},
  \bibinfo{pages}{035115} (\bibinfo{year}{2012}).

\bibitem[{\citenamefont{Aryasetiawan et~al.}(1996)\citenamefont{Aryasetiawan,
  Hedin, and Karlsson}}]{aryasetiawan}
\bibinfo{author}{\bibfnamefont{F.}~\bibnamefont{Aryasetiawan}},
  \bibinfo{author}{\bibfnamefont{L.}~\bibnamefont{Hedin}}, \bibnamefont{and}
  \bibinfo{author}{\bibfnamefont{K.}~\bibnamefont{Karlsson}},
  \bibinfo{journal}{Phys. Rev. Lett.} \textbf{\bibinfo{volume}{77}},
  \bibinfo{pages}{2268} (\bibinfo{year}{1996}).

\bibitem[{\citenamefont{Guzzo et~al.}(2011)\citenamefont{Guzzo, Lani, Sottile,
  Romaniello, Gatti, Kas, Rehr, Silly, Sirotti, and Reining}}]{guzzo}
\bibinfo{author}{\bibfnamefont{M.}~\bibnamefont{Guzzo}},
  \bibinfo{author}{\bibfnamefont{G.}~\bibnamefont{Lani}},
  \bibinfo{author}{\bibfnamefont{F.}~\bibnamefont{Sottile}},
  \bibinfo{author}{\bibfnamefont{P.}~\bibnamefont{Romaniello}},
  \bibinfo{author}{\bibfnamefont{M.}~\bibnamefont{Gatti}},
  \bibinfo{author}{\bibfnamefont{J.~J.} \bibnamefont{Kas}},
  \bibinfo{author}{\bibfnamefont{J.~J.} \bibnamefont{Rehr}},
  \bibinfo{author}{\bibfnamefont{M.~G.} \bibnamefont{Silly}},
  \bibinfo{author}{\bibfnamefont{F.}~\bibnamefont{Sirotti}}, \bibnamefont{and}
  \bibinfo{author}{\bibfnamefont{L.}~\bibnamefont{Reining}},
  \bibinfo{journal}{Phys. Rev. Lett.} \textbf{\bibinfo{volume}{107}},
  \bibinfo{pages}{166401} (\bibinfo{year}{2011}).

\bibitem[{\citenamefont{Zhou et~al.}(2015)\citenamefont{Zhou, Kas, Sponza,
  Reshetnyak, Guzzo, Giorgetti, Gatti, Sottile, Rehr, and Reining}}]{sky}
\bibinfo{author}{\bibfnamefont{J.}~\bibnamefont{Zhou}},
  \bibinfo{author}{\bibfnamefont{J.}~\bibnamefont{Kas}},
  \bibinfo{author}{\bibfnamefont{L.}~\bibnamefont{Sponza}},
  \bibinfo{author}{\bibfnamefont{I.}~\bibnamefont{Reshetnyak}},
  \bibinfo{author}{\bibfnamefont{M.}~\bibnamefont{Guzzo}},
  \bibinfo{author}{\bibfnamefont{C.}~\bibnamefont{Giorgetti}},
  \bibinfo{author}{\bibfnamefont{M.}~\bibnamefont{Gatti}},
  \bibinfo{author}{\bibfnamefont{F.}~\bibnamefont{Sottile}},
  \bibinfo{author}{\bibfnamefont{J.}~\bibnamefont{Rehr}}, \bibnamefont{and}
  \bibinfo{author}{\bibfnamefont{L.}~\bibnamefont{Reining}},
  \bibinfo{journal}{J. Chem. Phys.} \textbf{\bibinfo{volume}{143}}
  (\bibinfo{year}{2015}).

\bibitem[{\citenamefont{Lischner et~al.}(2013)\citenamefont{Lischner,
  Vigil-Fowler, and Louie}}]{lischner}
\bibinfo{author}{\bibfnamefont{J.}~\bibnamefont{Lischner}},
  \bibinfo{author}{\bibfnamefont{D.}~\bibnamefont{Vigil-Fowler}},
  \bibnamefont{and} \bibinfo{author}{\bibfnamefont{S.~G.} \bibnamefont{Louie}},
  \bibinfo{journal}{Phys. Rev. Lett.} \textbf{\bibinfo{volume}{110}},
  \bibinfo{pages}{146801} (\bibinfo{year}{2013}).

\bibitem[{\citenamefont{Caruso and Giustino}(2015)}]{giustino}
\bibinfo{author}{\bibfnamefont{F.}~\bibnamefont{Caruso}} \bibnamefont{and}
  \bibinfo{author}{\bibfnamefont{F.}~\bibnamefont{Giustino}},
  \bibinfo{journal}{Phys. Rev. B} \textbf{\bibinfo{volume}{92}},
  \bibinfo{pages}{045123} (\bibinfo{year}{2015}).

\bibitem[{\citenamefont{Rehr et~al.}(2009)\citenamefont{Rehr, Kas, Prange,
  Sorini, Takimoto, and Vila}}]{feffcr}
\bibinfo{author}{\bibfnamefont{J.~J.} \bibnamefont{Rehr}},
  \bibinfo{author}{\bibfnamefont{J.~J.} \bibnamefont{Kas}},
  \bibinfo{author}{\bibfnamefont{M.~P.} \bibnamefont{Prange}},
  \bibinfo{author}{\bibfnamefont{A.~P.} \bibnamefont{Sorini}},
  \bibinfo{author}{\bibfnamefont{Y.}~\bibnamefont{Takimoto}}, \bibnamefont{and}
  \bibinfo{author}{\bibfnamefont{F.}~\bibnamefont{Vila}}, \bibinfo{journal}{C.
  R. Phys.} \textbf{\bibinfo{volume}{10}} (\bibinfo{year}{2009}).

\bibitem[{\citenamefont{Calandra et~al.}(2012)\citenamefont{Calandra, Rueff,
  Gougoussis, C\'eolin, Gorgoi, Benedetti, Torelli, Shukla, Chandesris, and
  Brouder}}]{calandra2012}
\bibinfo{author}{\bibfnamefont{M.}~\bibnamefont{Calandra}},
  \bibinfo{author}{\bibfnamefont{J.~P.} \bibnamefont{Rueff}},
  \bibinfo{author}{\bibfnamefont{C.}~\bibnamefont{Gougoussis}},
  \bibinfo{author}{\bibfnamefont{D.}~\bibnamefont{C\'eolin}},
  \bibinfo{author}{\bibfnamefont{M.}~\bibnamefont{Gorgoi}},
  \bibinfo{author}{\bibfnamefont{S.}~\bibnamefont{Benedetti}},
  \bibinfo{author}{\bibfnamefont{P.}~\bibnamefont{Torelli}},
  \bibinfo{author}{\bibfnamefont{A.}~\bibnamefont{Shukla}},
  \bibinfo{author}{\bibfnamefont{D.}~\bibnamefont{Chandesris}},
  \bibnamefont{and} \bibinfo{author}{\bibfnamefont{C.}~\bibnamefont{Brouder}},
  \bibinfo{journal}{Phys. Rev. B} \textbf{\bibinfo{volume}{86}},
  \bibinfo{pages}{165102} (\bibinfo{year}{2012}).

\bibitem[{\citenamefont{Nozi\`{e}res and de~Dominicis}(1969)}]{ND}
\bibinfo{author}{\bibfnamefont{P.}~\bibnamefont{Nozi\`{e}res}}
  \bibnamefont{and} \bibinfo{author}{\bibfnamefont{C.~T.}
  \bibnamefont{de~Dominicis}}, \bibinfo{journal}{Phys. Rev.}
  \textbf{\bibinfo{volume}{178}}, \bibinfo{pages}{1097} (\bibinfo{year}{1969}).

\bibitem[{\citenamefont{Landau}(1944)}]{landau44}
\bibinfo{author}{\bibfnamefont{L.}~\bibnamefont{Landau}}, \bibinfo{journal}{J.
  Phys. USSR} \textbf{\bibinfo{volume}{8}}, \bibinfo{pages}{201}
  (\bibinfo{year}{1944}).

\bibitem[{\citenamefont{Bardyszewski and Hedin}(1985)}]{bard85}
\bibinfo{author}{\bibfnamefont{W.}~\bibnamefont{Bardyszewski}}
  \bibnamefont{and} \bibinfo{author}{\bibfnamefont{L.}~\bibnamefont{Hedin}},
  \bibinfo{journal}{Phys. Scr.} \textbf{\bibinfo{volume}{32}},
  \bibinfo{pages}{439} (\bibinfo{year}{1985}).

\bibitem[{\citenamefont{Kas et~al.}(2015)\citenamefont{Kas, Vila, Rehr, and
  Chambers}}]{KRC}
\bibinfo{author}{\bibfnamefont{J.~J.} \bibnamefont{Kas}},
  \bibinfo{author}{\bibfnamefont{F.~D.} \bibnamefont{Vila}},
  \bibinfo{author}{\bibfnamefont{J.~J.} \bibnamefont{Rehr}}, \bibnamefont{and}
  \bibinfo{author}{\bibfnamefont{S.~A.} \bibnamefont{Chambers}},
  \bibinfo{journal}{Phys. Rev. B} \textbf{\bibinfo{volume}{91}},
  \bibinfo{pages}{121112} (\bibinfo{year}{2015}).

\bibitem[{\citenamefont{Yabana and Bertsch}(1996)}]{yabana96}
\bibinfo{author}{\bibfnamefont{K.}~\bibnamefont{Yabana}} \bibnamefont{and}
  \bibinfo{author}{\bibfnamefont{G.~F.} \bibnamefont{Bertsch}},
  \bibinfo{journal}{Phys. Rev. B} \textbf{\bibinfo{volume}{54}},
  \bibinfo{pages}{4484} (\bibinfo{year}{1996}).

\bibitem[{\citenamefont{Yabana et~al.}(2006)\citenamefont{Yabana, Nakatsukasa,
  Iwata, and Bertsch}}]{yabana2006}
\bibinfo{author}{\bibfnamefont{K.}~\bibnamefont{Yabana}},
  \bibinfo{author}{\bibfnamefont{T.}~\bibnamefont{Nakatsukasa}},
  \bibinfo{author}{\bibfnamefont{J.-I.} \bibnamefont{Iwata}}, \bibnamefont{and}
  \bibinfo{author}{\bibfnamefont{G.~F.} \bibnamefont{Bertsch}},
  \bibinfo{journal}{Phys. Status Solidi B} \textbf{\bibinfo{volume}{243}},
  \bibinfo{pages}{1121} (\bibinfo{year}{2006}).

\bibitem[{\citenamefont{Takimoto et~al.}(2007)\citenamefont{Takimoto, Vila, and
  Rehr}}]{takimoto2007}
\bibinfo{author}{\bibfnamefont{Y.}~\bibnamefont{Takimoto}},
  \bibinfo{author}{\bibfnamefont{F.~D.} \bibnamefont{Vila}}, \bibnamefont{and}
  \bibinfo{author}{\bibfnamefont{J.~J.} \bibnamefont{Rehr}},
  \bibinfo{journal}{J. Chem. Phys.} \textbf{\bibinfo{volume}{127}},
  \bibinfo{pages}{154114} (\bibinfo{year}{2007}).

\bibitem[{\citenamefont{Vila et~al.}(2010)\citenamefont{Vila, Strubbe,
  Takimoto, Andrade, Rubio, Louie, and Rehr}}]{vila2010}
\bibinfo{author}{\bibfnamefont{F.~D.} \bibnamefont{Vila}},
  \bibinfo{author}{\bibfnamefont{D.~A.} \bibnamefont{Strubbe}},
  \bibinfo{author}{\bibfnamefont{Y.}~\bibnamefont{Takimoto}},
  \bibinfo{author}{\bibfnamefont{X.}~\bibnamefont{Andrade}},
  \bibinfo{author}{\bibfnamefont{A.}~\bibnamefont{Rubio}},
  \bibinfo{author}{\bibfnamefont{S.~G.} \bibnamefont{Louie}}, \bibnamefont{and}
  \bibinfo{author}{\bibfnamefont{J.~J.} \bibnamefont{Rehr}},
  \bibinfo{journal}{J. Chem. Phys.} \textbf{\bibinfo{volume}{133}},
  \bibinfo{pages}{034111} (\bibinfo{year}{2010}).

\bibitem[{\citenamefont{Otobe et~al.}(2009)\citenamefont{Otobe, Yabana, and
  Iwata}}]{otobe2009}
\bibinfo{author}{\bibfnamefont{T.}~\bibnamefont{Otobe}},
  \bibinfo{author}{\bibfnamefont{K.}~\bibnamefont{Yabana}}, \bibnamefont{and}
  \bibinfo{author}{\bibfnamefont{J.-I.} \bibnamefont{Iwata}},
  \bibinfo{journal}{J. Comput. Theory Nanosci} \textbf{\bibinfo{volume}{6}},
  \bibinfo{pages}{2545} (\bibinfo{year}{2009}).

\bibitem[{\citenamefont{Lee et~al.}(2012)\citenamefont{Lee, Vila, and
  Rehr}}]{ajlee}
\bibinfo{author}{\bibfnamefont{A.~J.} \bibnamefont{Lee}},
  \bibinfo{author}{\bibfnamefont{F.~D.} \bibnamefont{Vila}}, \bibnamefont{and}
  \bibinfo{author}{\bibfnamefont{J.~J.} \bibnamefont{Rehr}},
  \bibinfo{journal}{Phys. Rev. B} \textbf{\bibinfo{volume}{86}},
  \bibinfo{pages}{115107} (\bibinfo{year}{2012}).

\bibitem[{\citenamefont{Langreth}(1970)}]{langreth70}
\bibinfo{author}{\bibfnamefont{D.~C.} \bibnamefont{Langreth}},
  \bibinfo{journal}{Phys. Rev. B} \textbf{\bibinfo{volume}{1}},
  \bibinfo{pages}{471} (\bibinfo{year}{1970}).

\bibitem[{\citenamefont{Soler et~al.}(2002)\citenamefont{Soler, Artacho, Gale,
  Garcia, Junquera, Ordejon, and Sanchez-Portal}}]{siesta}
\bibinfo{author}{\bibfnamefont{J.}~\bibnamefont{Soler}},
  \bibinfo{author}{\bibfnamefont{E.}~\bibnamefont{Artacho}},
  \bibinfo{author}{\bibfnamefont{J.}~\bibnamefont{Gale}},
  \bibinfo{author}{\bibfnamefont{A.}~\bibnamefont{Garcia}},
  \bibinfo{author}{\bibfnamefont{J.}~\bibnamefont{Junquera}},
  \bibinfo{author}{\bibfnamefont{P.}~\bibnamefont{Ordejon}}, \bibnamefont{and}
  \bibinfo{author}{\bibfnamefont{D.}~\bibnamefont{Sanchez-Portal}},
  \bibinfo{journal}{J. Phys.: Condens. Matter} \textbf{\bibinfo{volume}{14}}
  (\bibinfo{year}{2002}).

\bibitem[{\citenamefont{Leiro et~al.}(2003)\citenamefont{Leiro, Heinonen,
  Laiho, and Batirev}}]{Leiro2003}
\bibinfo{author}{\bibfnamefont{J.}~\bibnamefont{Leiro}},
  \bibinfo{author}{\bibfnamefont{M.}~\bibnamefont{Heinonen}},
  \bibinfo{author}{\bibfnamefont{T.}~\bibnamefont{Laiho}}, \bibnamefont{and}
  \bibinfo{author}{\bibfnamefont{I.}~\bibnamefont{Batirev}},
  \bibinfo{journal}{J. Electron. Spectrosc. Relat. Phenom.}
  \textbf{\bibinfo{volume}{128}}, \bibinfo{pages}{205 } (\bibinfo{year}{2003}).

\bibitem[{\citenamefont{Shirley et~al.}()\citenamefont{Shirley, Fossard,
  Gilmore, Hug, Kas, Rehr, and Vila}}]{Shirleyetal16}
\bibinfo{author}{\bibfnamefont{E.~L.} \bibnamefont{Shirley}},
  \bibinfo{author}{\bibfnamefont{F.}~\bibnamefont{Fossard}},
  \bibinfo{author}{\bibfnamefont{K.}~\bibnamefont{Gilmore}},
  \bibinfo{author}{\bibfnamefont{G.}~\bibnamefont{Hug}},
  \bibinfo{author}{\bibfnamefont{J.~J.} \bibnamefont{Kas}},
  \bibinfo{author}{\bibfnamefont{J.~J.} \bibnamefont{Rehr}}, \bibnamefont{and}
  \bibinfo{author}{\bibfnamefont{F.~D.} \bibnamefont{Vila}},
  \bibinfo{note}{unpublished}.

\bibitem[{\citenamefont{Anderson}(1967)}]{anderson67}
\bibinfo{author}{\bibfnamefont{P.~W.} \bibnamefont{Anderson}},
  \bibinfo{journal}{Phys. Rev. Lett.} \textbf{\bibinfo{volume}{18}},
  \bibinfo{pages}{1049} (\bibinfo{year}{1967}).

\bibitem[{\citenamefont{Doniach and Sunjic}(1970)}]{doniach70}
\bibinfo{author}{\bibfnamefont{S.}~\bibnamefont{Doniach}} \bibnamefont{and}
  \bibinfo{author}{\bibfnamefont{M.}~\bibnamefont{Sunjic}},
  \bibinfo{journal}{J. Phys. C (Solid State)} \textbf{\bibinfo{volume}{3}},
  \bibinfo{pages}{285} (\bibinfo{year}{1970}).

\bibitem[{\citenamefont{Goubin et~al.}(2004)\citenamefont{Goubin, Rocquefelte,
  Whangbo, Montardi, Brec, and Jobic}}]{CeO2Loss}
\bibinfo{author}{\bibfnamefont{F.}~\bibnamefont{Goubin}},
  \bibinfo{author}{\bibfnamefont{X.}~\bibnamefont{Rocquefelte}},
  \bibinfo{author}{\bibfnamefont{M.-H.} \bibnamefont{Whangbo}},
  \bibinfo{author}{\bibfnamefont{Y.}~\bibnamefont{Montardi}},
  \bibinfo{author}{\bibfnamefont{R.}~\bibnamefont{Brec}}, \bibnamefont{and}
  \bibinfo{author}{\bibfnamefont{S.}~\bibnamefont{Jobic}},
  \bibinfo{journal}{Chem. Mater.} \textbf{\bibinfo{volume}{16}},
  \bibinfo{pages}{662} (\bibinfo{year}{2004}).

\bibitem[{\citenamefont{Kas et~al.}(2014)\citenamefont{Kas, Rehr, and
  Reining}}]{KasRC}
\bibinfo{author}{\bibfnamefont{J.~J.} \bibnamefont{Kas}},
  \bibinfo{author}{\bibfnamefont{J.~J.} \bibnamefont{Rehr}}, \bibnamefont{and}
  \bibinfo{author}{\bibfnamefont{L.}~\bibnamefont{Reining}},
  \bibinfo{journal}{Phys. Rev. B} \textbf{\bibinfo{volume}{90}},
  \bibinfo{pages}{085112} (\bibinfo{year}{2014}).

\bibitem[{\citenamefont{Lawler et~al.}(2008)\citenamefont{Lawler, Rehr, Vila,
  Dalosto, Shirley, and Levine}}]{AI2NBSE}
\bibinfo{author}{\bibfnamefont{H.~M.} \bibnamefont{Lawler}},
  \bibinfo{author}{\bibfnamefont{J.~J.} \bibnamefont{Rehr}},
  \bibinfo{author}{\bibfnamefont{F.}~\bibnamefont{Vila}},
  \bibinfo{author}{\bibfnamefont{S.~D.} \bibnamefont{Dalosto}},
  \bibinfo{author}{\bibfnamefont{E.~L.} \bibnamefont{Shirley}},
  \bibnamefont{and} \bibinfo{author}{\bibfnamefont{Z.~H.}
  \bibnamefont{Levine}}, \bibinfo{journal}{Phys. Rev. B}
  \textbf{\bibinfo{volume}{78}}, \bibinfo{pages}{205108}
  (\bibinfo{year}{2008}).

\bibitem[{\citenamefont{Sokcevic et~al.}(1979)\citenamefont{Sokcevic, Sunjik,
  and Fadley}}]{sokcevik79}
\bibinfo{author}{\bibfnamefont{D.}~\bibnamefont{Sokcevic}},
  \bibinfo{author}{\bibfnamefont{M.}~\bibnamefont{Sunjik}}, \bibnamefont{and}
  \bibinfo{author}{\bibfnamefont{C.}~\bibnamefont{Fadley}},
  \bibinfo{journal}{Surf. Sci.} \textbf{\bibinfo{volume}{82}},
  \bibinfo{pages}{383} (\bibinfo{year}{1979}).

\bibitem[{\citenamefont{Hedin}(1980)}]{hedinrecoil80}
\bibinfo{author}{\bibfnamefont{L.}~\bibnamefont{Hedin}},
  \bibinfo{journal}{Phys. Scr.} \textbf{\bibinfo{volume}{21}},
  \bibinfo{pages}{477} (\bibinfo{year}{1980}).

\bibitem[{\citenamefont{Klevak et~al.}(2014)\citenamefont{Klevak, Kas, and
  Rehr}}]{klevak2014}
\bibinfo{author}{\bibfnamefont{E.}~\bibnamefont{Klevak}},
  \bibinfo{author}{\bibfnamefont{J.~J.} \bibnamefont{Kas}}, \bibnamefont{and}
  \bibinfo{author}{\bibfnamefont{J.~J.} \bibnamefont{Rehr}},
  \bibinfo{journal}{Phys. Rev. B} \textbf{\bibinfo{volume}{89}},
  \bibinfo{pages}{085123} (\bibinfo{year}{2014}).

\bibitem[{\citenamefont{Lagarde}(1999)}]{lagarde99}
\bibinfo{author}{\bibfnamefont{P.}~\bibnamefont{Lagarde}},
  \bibinfo{howpublished}{private communication} (\bibinfo{year}{1999}).

\bibitem[{\citenamefont{Karis et~al.}(2008)\citenamefont{Karis, Svensson, Rusz,
  Oppeneer, Gorgoi, Sch\"afers, Braun, Eberhardt, and
  M\aa{}rtensson}}]{karis2008}
\bibinfo{author}{\bibfnamefont{O.}~\bibnamefont{Karis}},
  \bibinfo{author}{\bibfnamefont{S.}~\bibnamefont{Svensson}},
  \bibinfo{author}{\bibfnamefont{J.}~\bibnamefont{Rusz}},
  \bibinfo{author}{\bibfnamefont{P.~M.} \bibnamefont{Oppeneer}},
  \bibinfo{author}{\bibfnamefont{M.}~\bibnamefont{Gorgoi}},
  \bibinfo{author}{\bibfnamefont{F.}~\bibnamefont{Sch\"afers}},
  \bibinfo{author}{\bibfnamefont{W.}~\bibnamefont{Braun}},
  \bibinfo{author}{\bibfnamefont{W.}~\bibnamefont{Eberhardt}},
  \bibnamefont{and}
  \bibinfo{author}{\bibfnamefont{N.}~\bibnamefont{M\aa{}rtensson}},
  \bibinfo{journal}{Phys. Rev. B} \textbf{\bibinfo{volume}{78}},
  \bibinfo{pages}{233105} (\bibinfo{year}{2008}).

\bibitem[{\citenamefont{Xu et~al.}(1995)\citenamefont{Xu, Liu, Johnson,
  Itchkawitz, Randall, Feldhaus, and Bradshaw}}]{xu1995}
\bibinfo{author}{\bibfnamefont{Z.}~\bibnamefont{Xu}},
  \bibinfo{author}{\bibfnamefont{Y.}~\bibnamefont{Liu}},
  \bibinfo{author}{\bibfnamefont{P.~D.} \bibnamefont{Johnson}},
  \bibinfo{author}{\bibfnamefont{B.}~\bibnamefont{Itchkawitz}},
  \bibinfo{author}{\bibfnamefont{K.}~\bibnamefont{Randall}},
  \bibinfo{author}{\bibfnamefont{J.}~\bibnamefont{Feldhaus}}, \bibnamefont{and}
  \bibinfo{author}{\bibfnamefont{A.}~\bibnamefont{Bradshaw}},
  \bibinfo{journal}{Phys. Rev. B} \textbf{\bibinfo{volume}{51}},
  \bibinfo{pages}{7912} (\bibinfo{year}{1995}).

\bibitem[{\citenamefont{Chen et~al.}(1991)\citenamefont{Chen, Smith, and
  Sette}}]{chen1991}
\bibinfo{author}{\bibfnamefont{C.~T.} \bibnamefont{Chen}},
  \bibinfo{author}{\bibfnamefont{N.~V.} \bibnamefont{Smith}}, \bibnamefont{and}
  \bibinfo{author}{\bibfnamefont{F.}~\bibnamefont{Sette}},
  \bibinfo{journal}{Phys. Rev. B} \textbf{\bibinfo{volume}{43}},
  \bibinfo{pages}{6785} (\bibinfo{year}{1991}).

\bibitem[{\citenamefont{Bagus et~al.}(2010{\natexlab{b}})\citenamefont{Bagus,
  Nelin, Ilton, Baron, Abbott, Primorac, Kuhlenbeck, Shaikhutdinov, and
  Freund}}]{CeO2XPS}
\bibinfo{author}{\bibfnamefont{P.}~\bibnamefont{Bagus}},
  \bibinfo{author}{\bibfnamefont{C.}~\bibnamefont{Nelin}},
  \bibinfo{author}{\bibfnamefont{E.}~\bibnamefont{Ilton}},
  \bibinfo{author}{\bibfnamefont{M.}~\bibnamefont{Baron}},
  \bibinfo{author}{\bibfnamefont{H.}~\bibnamefont{Abbott}},
  \bibinfo{author}{\bibfnamefont{E.}~\bibnamefont{Primorac}},
  \bibinfo{author}{\bibfnamefont{H.}~\bibnamefont{Kuhlenbeck}},
  \bibinfo{author}{\bibfnamefont{S.}~\bibnamefont{Shaikhutdinov}},
  \bibnamefont{and} \bibinfo{author}{\bibfnamefont{H.-J.}
  \bibnamefont{Freund}}, \bibinfo{journal}{Chem. Phys. Lett.}
  \textbf{\bibinfo{volume}{487}}, \bibinfo{pages}{237 }
  (\bibinfo{year}{2010}{\natexlab{b}}).

\bibitem[{\citenamefont{Kotani et~al.}(2012)\citenamefont{Kotani, Kvashnina,
  Butorin, and Glatzel}}]{kotani2012}
\bibinfo{author}{\bibfnamefont{A.}~\bibnamefont{Kotani}},
  \bibinfo{author}{\bibfnamefont{K.~O.} \bibnamefont{Kvashnina}},
  \bibinfo{author}{\bibfnamefont{S.~M.} \bibnamefont{Butorin}},
  \bibnamefont{and} \bibinfo{author}{\bibfnamefont{P.}~\bibnamefont{Glatzel}},
  \bibinfo{journal}{EPJ B} \textbf{\bibinfo{volume}{85}}, \bibinfo{pages}{1}
  (\bibinfo{year}{2012}).

\bibitem[{\citenamefont{Lee et~al.}(1999)\citenamefont{Lee, Gunnarsson, and
  Hedin}}]{lgh}
\bibinfo{author}{\bibfnamefont{J.~D.} \bibnamefont{Lee}},
  \bibinfo{author}{\bibfnamefont{O.}~\bibnamefont{Gunnarsson}},
  \bibnamefont{and} \bibinfo{author}{\bibfnamefont{L.}~\bibnamefont{Hedin}},
  \bibinfo{journal}{Phys. Rev. B} \textbf{\bibinfo{volume}{60}},
  \bibinfo{pages}{8034} (\bibinfo{year}{1999}).

\bibitem[{\citenamefont{Vila et~al.}(2007)\citenamefont{Vila, Rehr, Rossner,
  and Krappe}}]{dmdw}
\bibinfo{author}{\bibfnamefont{F.~D.} \bibnamefont{Vila}},
  \bibinfo{author}{\bibfnamefont{J.~J.} \bibnamefont{Rehr}},
  \bibinfo{author}{\bibfnamefont{H.~H.} \bibnamefont{Rossner}},
  \bibnamefont{and} \bibinfo{author}{\bibfnamefont{H.~J.}
  \bibnamefont{Krappe}}, \bibinfo{journal}{Phys. Rev. B}
  \textbf{\bibinfo{volume}{76}}, \bibinfo{pages}{014301}
  (\bibinfo{year}{2007}).

\bibitem[{\citenamefont{Story et~al.}(2014)\citenamefont{Story, Kas, Vila,
  Verstraete, and Rehr}}]{story2014}
\bibinfo{author}{\bibfnamefont{S.~M.} \bibnamefont{Story}},
  \bibinfo{author}{\bibfnamefont{J.~J.} \bibnamefont{Kas}},
  \bibinfo{author}{\bibfnamefont{F.~D.} \bibnamefont{Vila}},
  \bibinfo{author}{\bibfnamefont{M.~J.} \bibnamefont{Verstraete}},
  \bibnamefont{and} \bibinfo{author}{\bibfnamefont{J.~J.} \bibnamefont{Rehr}},
  \bibinfo{journal}{Phys. Rev. B} \textbf{\bibinfo{volume}{90}},
  \bibinfo{pages}{195135} (\bibinfo{year}{2014}).

\bibitem[{\citenamefont{Gunnarsson et~al.}(1994)\citenamefont{Gunnarsson,
  Meden, and Sch\"onhammer}}]{gunnarsson1994}
\bibinfo{author}{\bibfnamefont{O.}~\bibnamefont{Gunnarsson}},
  \bibinfo{author}{\bibfnamefont{V.}~\bibnamefont{Meden}}, \bibnamefont{and}
  \bibinfo{author}{\bibfnamefont{K.}~\bibnamefont{Sch\"onhammer}},
  \bibinfo{journal}{Phys. Rev. B} \textbf{\bibinfo{volume}{50}},
  \bibinfo{pages}{10462} (\bibinfo{year}{1994}).

\end{thebibliography}
\end{document}